 \documentclass[acmtog,authorversion]{acmart}
\acmSubmissionID{307}

\AtBeginDocument{%
  \providecommand\BibTeX{{%
    \normalfont B\kern-0.5em{\scshape i\kern-0.25em b}\kern-0.8em\TeX}}}

\setcopyright{acmcopyright}

\acmJournal{TOG}
\acmYear{2020}
\acmVolume{39}
\acmNumber{4}
\acmArticle{1}
\acmMonth{7}
\acmDOI{10.1145/3386569.3392424}

\usepackage{enumitem}
\usepackage{subcaption}

\newcommand{\noarxiv}[1]{}

\usepackage{accents}
\newlength{\dhatheight}

\usepackage{mathtools}

\usepackage{xspace}
\makeatletter
\DeclareRobustCommand\onedot{\futurelet\@let@token\@onedot}
\def\@onedot{\ifx\@let@token.\else.\null\fi\xspace}

\def\eg{{e.g}\onedot} 
\def\ie{{i.e}\onedot}

\makeatother

\usepackage{sistyle}
\SIthousandsep{,}

\usepackage{upgreek}

\def\1n{\mathbf{1}_n}
\def\0{\mathbf{0}}
\def\1{\mathbf{1}}

\def\R{{\mathbb R}}

\def\n{{\bf n}}

\def\bphi{\mbox{\boldmath{$\phi$}}}

\newcommand{\bu}{\mathbf{u}}

\newcommand{\ms}{ms\xspace}

\newcommand{\cm}{cm\xspace}
\newcommand{\mm}{mm\xspace}
\newcommand{\nm}{nm\xspace}
\newcommand{\um}{$\upmu$m\xspace}

 \usepackage{physics}

\def\BE{\begin{equation}}
\def\EE{\end{equation}}
\def\BEA{\begin{eqnarray}}
\def\EEA{\end{eqnarray}}

\newcommand{\edit}[1]{#1} %
\newcommand{\newedit}[1]{#1}  %

\newcommand{\secsym}{Sec\onedot}
\newcommand{\figsym}{Fig\onedot}
\newcommand{\equsym}{Eq\onedot}

\newcommand{\figsymp}{Figs\onedot}

\newcommand{\secref}[1]{\secsym~\ref{#1}}
\newcommand{\figref}[1]{\figsym~\ref{#1}}

\newcommand{\conetilt}{ConeTilt\xspace}

\usepackage{placeins}

\begin{document}

\title{Towards Occlusion-Aware Multifocal Displays}

\author{Jen-Hao Rick Chang}
\orcid{1234-5678-9012-3456}
\affiliation{%
	\institution{Carnegie Mellon University}
	\streetaddress{5000 Forbes Ave}
	\city{Pittsburgh}
	\state{PA}
	\postcode{15213}
	\country{USA}}
\email{rickchang@cmu.edu}
\author{Anat Levin}
\affiliation{%
	\institution{Technion}
	\city{Haifa}
	\country{Israel}
}
\email{anat.levin@ee.technion.ac.il}
\author{B.\ V.\ K.\ Vijaya Kumar}
\affiliation{%
	\institution{Carnegie Mellon University}
	\city{Pittsburgh}
	\country{USA}
}
\email{kumar@ece.cmu.edu}
\author{Aswin C.\ Sankaranarayanan}
\affiliation{%
	\institution{Carnegie Mellon University}
	\city{Pittsburgh}
	\country{USA}
}
\email{saswin@andrew.cmu.edu}

\renewcommand{\shortauthors}{Chang et al.}

\begin{abstract}
The human visual system uses numerous cues for depth perception, including disparity, accommodation, motion parallax and occlusion.
It is incumbent upon virtual-reality displays to satisfy these cues to provide an immersive user experience.
Multifocal displays, one of the classic approaches to satisfy the accommodation cue, place virtual content at multiple focal planes, each at a different depth.
However, the content on focal planes close to the eye do not occlude those farther away; this deteriorates the occlusion cue as well as reduces contrast at depth discontinuities due to leakage of the defocus blur.
This paper enables occlusion-aware multifocal displays using a novel \conetilt operator that provides an additional degree of freedom --- tilting the light cone emitted at each pixel of the display panel.
We show that, for  scenes with relatively simple occlusion configurations, tilting the light cones provides the same effect as physical occlusion.
We demonstrate that \conetilt can be easily implemented by a phase-only spatial light modulator.
Using a lab prototype, we show results that demonstrate the presence of occlusion cues and the increased contrast of the display at depth edges.
\end{abstract}



\begin{teaserfigure}
	\includegraphics[width=\linewidth]{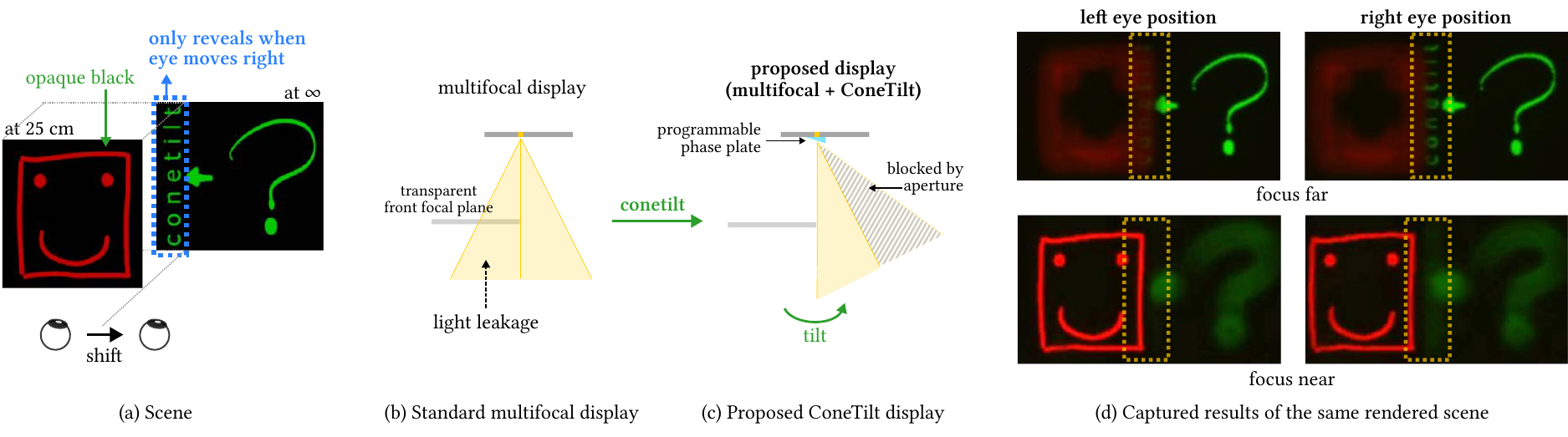}
	\vspace{-7mm}
	\caption{
		Multifocal displays  present content at multiple focal planes to satisfy the accommodation cue of the human vision. (a) A scene with content on two focal planes. The text ``conetilt'' is expected to be occluded when the eye shifts laterally. (b) However, in a  standard multifocal display, content on focal planes do not occlude each other and  hence content from a farther plane can leak into the area of occluding objects at frontal planes. (c) We propose a \conetilt multifocal display which  provides the ability to tilt the cone of light emerging from each pixel and hence produce the same effect as occlusion. (d) The images captured by our prototype for the scene in (a) under \edit{$\pm0.5$ \mm} lateral translation of a camera. Note how the text next to the boundary is hidden on one viewpoint and revealed on the other, and how defocus cues are faithfully reproduced. This result was produced without gaze tracking or content re-rendering.
	}
\label{fig:teaser}
\end{teaserfigure}

\maketitle

\section{Introduction} \label{sec:intro}

The primary aim of a virtual-reality (VR) display is to present a scene to the eye that is indistinguishable from reality.
In the context of depth perception, a VR display has to faithfully reproduce the visual cues pertaining to disparity, accommodation, occlusion, and motion parallax.
While there are many types of VR displays, each with differing amounts of fidelity towards satisfying these cues, this paper focuses on multifocal displays with the objective of enhancing the range of perceptual cues that they can satisfy.

In a multifocal display, three-dimensional (3D) content is shown to a user by placing virtual objects on different focal planes, which are optically placed at different depths from the viewer.
This has a unique advantage that the display automatically renders the accommodation cues, \ie, supports the focus of our eyes, provided there are a sufficient number of focal planes \cite{rolland1999dynamic,mackenzie2010accommodation,watt2012real,chang2018towards}. %
In order to display multiple focal planes at different depths, the focal planes are time-multiplexed and  content on the planes does not occlude  each other.
Hence, the focal planes behave as if they were transparent, as seen in \figsym\ref{fig:teaser}b.
This inability to occluded light leads to two adverse effects.
First, the display is incapable of satisfying occlusion cues since even small displacements of the eye will readily produce overlapping contents. %
Second, even when the eye is positioned correctly, the contrast of depth edges is greatly reduced.
This can be seen from the example shown in \figsym~\ref{figure: sunny}c; when our eyes focus on the dinosaur and the far focal planes get defocused, the defocused background bleeds into the dinosaur and reduces its contrast.
Both of these effects reduce the immersive nature of the VR experience.

One potential approach for enabling occlusion cues on a multifocal display is to use a light-field display.
The improved angular resolution allows us to control the intensity of the light rays that a pixel sends in different directions.
The occlusion cue can then be produced by avoiding sending light through any virtual opaque object on the front focal planes. 
However, the additional angular resolution usually comes at the cost of significant loss in  spatial resolution~\cite{lanman2013near,huang2015light}.
The loss of contrast at depth edges can also potentially be addressed by optimizing the content shown on focal planes to account for the transparency \cite{narain2015optimal,padmanaban2017optimizing}.
However, such optimization is tuned to a specific view point and the visual immersion breaks down when the viewpoint is shifted even slightly.

This paper provides a design for multifocal displays, capable of rendering occlusion cues, without any loss of spatial resolution. %
Our key idea is that to satisfy occlusion cues, for most scenes we do not need real angular resolution in the physical display, but simply the ability to \textit{tilt the light cone} emitted by display pixels. 
With appropriate tilts of the light cones, we can emulate the same effect as physical occlusion between real objects.

\figsym \ref{fig:teaser}c illustrates the case when we try to partially occlude a pixel on a far focal plane. 
For pixels on the far focal plane, since the occluder is on the left, tilting the light cone emitted by the pixel to the right will ensure that no light rays from the pixel pass through the occluder and thereby creates an illusion that the front occluder blocks light.
As a result, the near content  successfully occludes far content; this is seen in the occlusion of the back content in \figsym~\ref{fig:teaser}d as well as realistic defocus blur at depth edges in  \figsym~\ref{figure: sunny}d.
More importantly, since the \textit{entire} light cone is tilted, we do not need additional angular resolution on the display panel and there is no loss of the spatial resolution of the display.

The tilt of the light cones emitted by display pixels is implemented by placing a phase-only spatial light modulator (phase SLM) on the display panel.  
By programming the slope of the phase function at each display pixel, we can steer the light cone emitted by each pixel.
The phase SLM acts as a \textit{freeform field lens} that dynamically tilts each light cone based on the virtual scene.

\subsection{Contributions}
We make the following contributions.
\begin{itemize}[leftmargin=*]
        \item \textit{ConeTilt multifocal displays.} Our primary contribution is the use of the \conetilt operation to endow occlusion cues in multifocal displays without loss of spatial resolution.
        \item \textit{Implementation.} We provide a simple approach for implementing \conetilt using phase SLMs. Given a virtual scene to be displayed, we derive the phase function operating the SLM. %
        \item \textit{Design space analysis.} We derive important properties of the \conetilt display including the fidelity of its occlusion cues, the contrast of the display, as well as the field-of-view and the size of the eye box.
        \item \textit{Prototype.} We build a lab prototype using off-the-shelf components and demonstrate a \conetilt  display in practice. 
\end{itemize}

\subsection{Limitations}
The proposed approach comes with the following limitations.
\begin{itemize}[leftmargin=*]
		\item \textit{Dark halo.} 
		The \conetilt operation is not completely equivalent to the cropped light cones achieved by physical occluders; this results in some dark halo artifacts next to depth   discontinuities.   
        \item \textit{Complexity of the occluding contours.} 
        While tilting the cone is sufficient for simple  occlusion boundaries, it is insufficient for more complex dense occlusions. 
        Examples include the occluding object being a mesh or dense foliage.
        \item \textit{Limitation of the prototype.}  Our prototype uses a phase SLM to implement the \conetilt operation and, as such, we are limited by its  ability to tilt light cones.
        Our prototype uses a phase SLM with $6.4$ \um pixel pitch, which can tilt the light cones up to $2^\circ$ in each direction and this places additional restrictions on our prototype.
\end{itemize}
We discuss these limitations in detail in \secref{sec: limitation}.

\begin{figure}
	\centering
	\includegraphics[width=\linewidth]{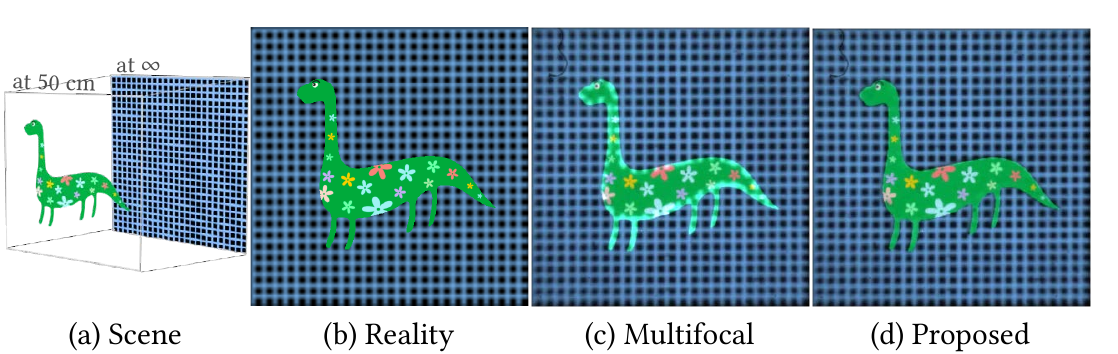}
	\caption{\textbf{Lack of occlusion cue and lowered contrast in multifocal displays.}  (a) A 3D scene  with content on two focal planes. We show real captured results on a lab prototype of (b) a multifocal display and (c) the proposed \conetilt display. In both cases, the observer is focused on the front plane.  The image captured on the multifocal display in (b) leaks background content into the front plane due to the inherent transparency of the focal planes.  In contrast, the image captured on the \conetilt display in (c) has crisp occlusion boundaries with no light leakage.}
	\label{figure: sunny}
\end{figure}

\section{Prior Work} \label{sec:prior}

We briefly discuss related research on  occlusion cues in VR displays.

\subsection{Role of Occlusion  in Visual Perception}

Among the  cues deployed by the human visual system to perceive the world, occlusion plays a dominant role~\cite{cutting1995perceiving,geng2013three}.
When two opaque objects are at different depths, the object in front will occlude some light rays from the object behind.
Moving our head and changing our perspective, even by a small amount, will reveal parts of the back object that was originally hidden.
The occluding and revealing of objects allows us to easily discover their relative depths even when the objects are in close proximity.
Further, when our eye focuses on objects at different depths, the subtle differences in the defocus blur at depth discontinuities are often sufficient to resolve their relative ordering \cite{zannoli2016blur}.
This makes occlusion one of the dominant cues for depth perception that works reliably across a wide depth range~\cite{cutting1995perceiving}.
As a consequence, it is of utmost importance that 3D displays, such as VR displays, generate occlusion cues properly.

\subsection{Enabling Occlusion Cues  in VR Displays}
Most commercial VR displays generate occlusion cues by tracking the head/eye and regenerating content from the new perspective. 
This ensures that occlusion cues are faithfully produced and is only limited by the refresh rate of the display.
However, as is often the case, the content is shown on a single plane and hence, there are gross accommodation errors.
To alleviate the problem, gaze trackers can be used to estimate the user's gaze and pupil position, and subsequently update the displayed content.
This increases both the hardware requirement and the computational cost.
We instead focus on enabling  displays that simultaneously produce the accommodation  and the occlusion cues under changing focus and  movements of the eyes.

There are many display technologies that can produce occlusion cues without tracking.
Cossairt et al.\ [\citeyear{Cossairt:07}] and Jones et al.\ [\citeyear{jones2007rendering}] produce volumetric displays by rotating an anisotropic diffuser in synchrony with a projector.
As the diffuser spins, the projector displays an image to be seen by a viewer in a specific direction.
This results in realizing occlusion without knowing the position of the viewer.
However, the spinning diffuser makes the displays more geared towards 3D televisions and not VR.

Light field displays \cite{lanman2013near,huang2014eyeglasses} provide angular control and, in principle, this is sufficient to produce rich occlusion cues.
However, the gain in angular resolution is invariably accompanied by a loss in spatial resolution of the display.
Further, the finite pixel pitch of the display greatly limits the depth range the displays can support, \ie, only content whose depth is in the vicinity of the display depth can be faithfully rendered.
While there are alternate implementations of light field displays that do not rely on microlens arrays \cite{huang2015light,wetzstein2011layered}, these do share the same challenges in obtaining a large depth range.
In comparison, the depth range of multifocal displays is determined by the focus tunable lens and is often more than several diopters.

The importance of occlusion cues and methods to achieve it have been studied extensively in the context of augmented reality (AR) displays \cite{kiyokawa2000optical,mulder2005realistic,inami2000visuo,rathinavel2019varifocal}.
However, these approaches concentrate on blocking light from real objects, wherein the challenges are different from those in VR displays.

\subsection{Multifocal Displays}
Multifocal displays \cite{akeley2004achieving} show content at multiple focal planes placed at different depths from the viewer.
The displays are capable of producing natural accommodation cues over a wide depth range~\edit{~\cite{koulieris2017accommodation}}.
There are many possible ways of implementing such a display including 
using a focus-tunable lens %
\cite{liu2008optical,liu2009time,love2009high,llull2015design,johnson2016dynamic,konrad2016novel,rathinavel2018extended,chang2018towards,lee2019tomographic,youngjin2019tomographic},  a waveplate lens~\cite{tabiryan2015thin}, or variable-focus Moir{\'e} lenses~\cite{bernet2008adjustable}.
Despite the wide varieties in the implementations, the content on different focal planes do not occlude each other and hence, the displays produce inconsistent occlusion  and defocus cues at depth discontinuities.
\edit{Please refer to \cite{koulieris2019near} for a recent survey on other AR/VR displays.}

\subsection{Other Related Methods}
\edit{
Phase SLMs have been used in many other works to manipulate light.
For example, \citeN{matsuda2017focal} use a phase SLM to create smooth focal surfaces to support natural vision accommodation in virtual-reality displays.
\citeN{maimone2017holographic} create hologram for virtual/augmented reality displays.
\citeN{levin2016passive} use two phase SLMs to create a passive viewpoint-sensitive display.
\citeN{damberg2016high} create goal-based caustics with a phase SLM to increase the contrast of a high-dynamic-range projector.
Note that our use of a phase SLM --- attaching it directly on the display panel to create a freeform field lens ---  is different from all the aforementioned works.
This enables us to tilt the light cones while retaining the spatial resolution of the display.
}

\begin{figure*}[th]
	\includegraphics[width=\textwidth]{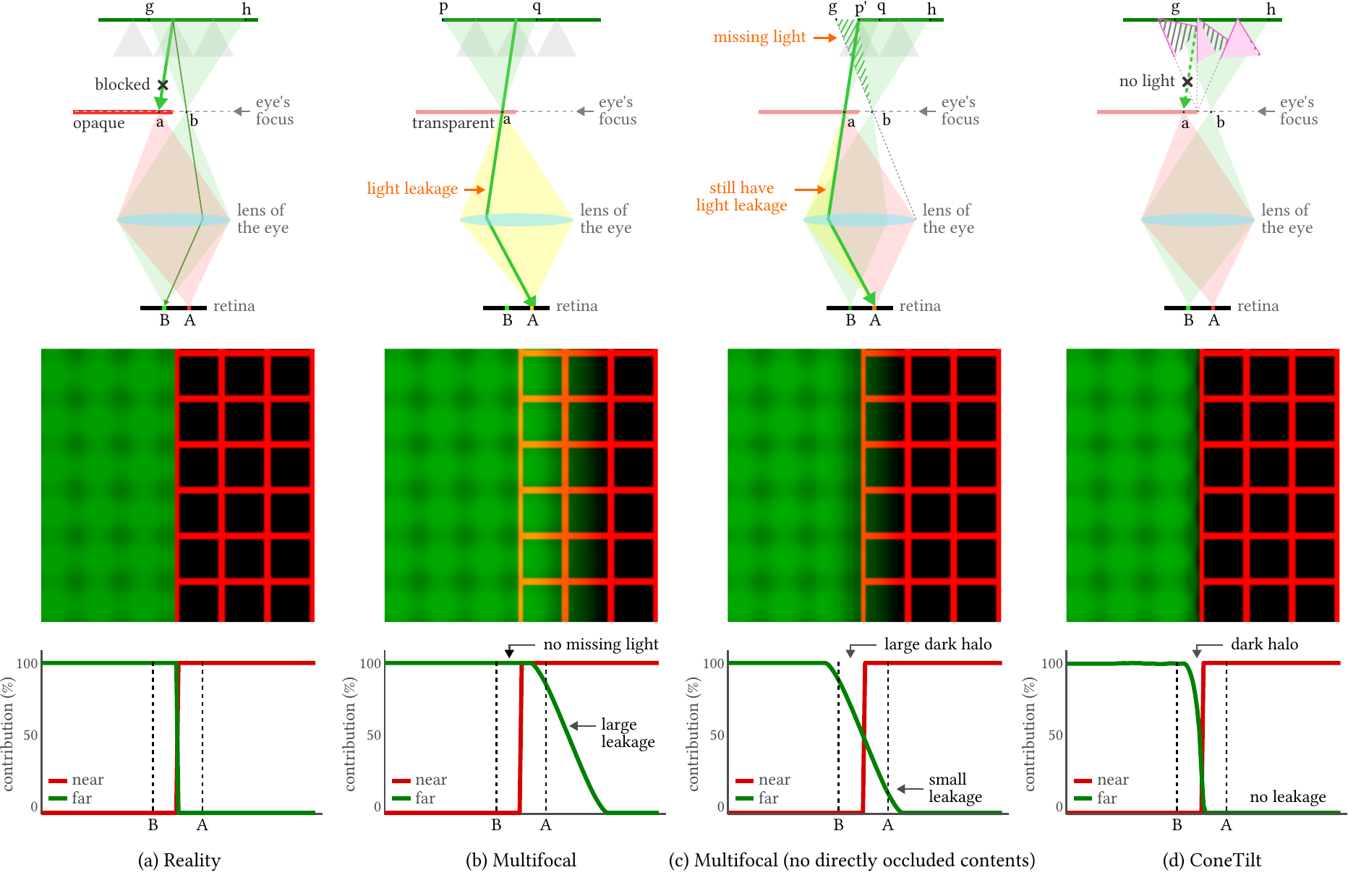}
	\caption{\textbf{The concept of \conetilt.} 
		We consider a scene consisting of two planes at different depths and show the image formation in (a) the real world, (b, c) multifocal displays with/without showing the overlapped part of the back plane, as well as (d) the proposed \conetilt displays. 
		In each case, the middle row shows a rendered image obtained when a camera/eye is focused on the front plane, and the bottom row shows the contribution from the front and the back planes (assuming all pixel values equal to 1).
		(a) In the real world, the front plane blocks the light from the back plane, and thus we see a sharp edge with no light from the back plane leaking onto the front. 
		(b, c) In a multifocal display, the inherent transparent nature of focal planes leads to light leakage from the back focal plane. The light leakage cannot be prevented even when we remove the overlapped region from the content shown on the back plane.
		(d) In a \conetilt display, the light cones are tilted to avoid emitting light rays that intersect with the content on the front focal plane, and thereby the display produces occlusion cues similar to those found in the real world. No light from the back plane leaks to the front plane, even when we do not remove the overlapping contents. Note that some light is missing from the back layer and creates a ``dark halo'' that we  explain in \secsym~\ref{sec: properties}.
	}
	\label{fig:idea}
\end{figure*}

\section{ConeTilt Multifocal Displays} \label{sec:coneshift}
We start by studying the occlusion cues in the real world and what happens in its absence in a multifocal display. 
Subsequently, we introduce the concept of \conetilt for producing occlusion cues.

\subsection{Occlusion Cues in Real Scenes}
Consider a scene consisting of two fronto-parallel planes, that are opaque and placed at different depths, as shown in \figref{fig:idea}a.
The front plane is red and the back plane is green; the camera/eye focuses on the front plane.
Consider two points $a$ and $b$, that are on either side of a depth discontinuity.
At point $a$, all the light coming from the back plane is blocked, due to the opaqueness of the front plane.
At point $b$, we get light from region $\overline{gh}$ on the back plane.
Since the camera focuses on the front plane, light passing through point $a$ and $b$ will be collected by pixel $A$ and $B$, respectively.
Since no green light from the back object passes through $a$, pixel $A$ is pure red.

\subsection{Occlusion Cues in Multifocal Displays}
Let us now consider the same scene, but rendered by a multifocal display.
For simplicity, we will assume that the two planes are displayed on focal planes corresponding to their true depth.
As with most multifocal designs, the focal planes are transparent, and as a result, light from the back focal plane can leak through the content shown on the front focal plane.
In \figref{fig:idea}b, pixel $A$ receives not only light emitted by point $a$ but also all the light from $\overline{pq}$ passing through $a$, making $A$ a yellow pixel (instead of red).  

The light leakage has two consequences.  
\begin{itemize}[leftmargin=*]
\item \textit{Loss of occlusion cue.} When two focal planes are in the depth of field of our eye, their contents will overlap even when we want to display an opaque front object.
\item \textit{Reduced contrast  ratio.} When we focus on the front plane (and the back focal plane is defocused), the front focal plane will be overlaid with the blurred content from behind and thereby lose its contrast.
The low contrast makes displaying dark objects on the front focal plane very difficult.
\end{itemize}

Removing occluded contents on the back focal plane cannot solve the leakage problem entirely.
In \figref{fig:idea}c, we remove the region behind the front object given the position of the eye; however, since each display pixel emits light toward a wide range of angles, light from the region $\overline{p'q}$ still leaks through point $a$ and reduces the contrast of pixel $A$.  
Removing occluded contents has another side effect --- it decreases the intensity of defocused content near depth discontinuities.
Let us use point $b$ as an example.  
In reality, point $b$ receives light from region $\overline{gh}$. 
Since we remove occluded region $\overline{gp'}$, we reduce the amount of light passing through point $b$ and thereby make pixel $B$ dimmer than the reality.

\subsection{Enabling Occlusion Cues via \conetilt }
The proposed display aims to produce occlusion cues on multifocal displays via a simple operation, that we refer to as \textit{\conetilt}. This allows  for the cone of light emanated at each display pixel to be \textit{independently tilted.}
We discuss the basic idea of \conetilt here and defer the details of its  implementation  to \secsym~\ref{sec:design}.

We consider the same scenario of a scene with two planes rendered by a multifocal display.
However, on the back focal plane, we apply a \conetilt operation at pixels near the occluding edge that is defined as follows: \textit{for each pixel, we tilt the cone such that no emitted light ray intersects with the content shown on the front plane.}
As is to be expected, the resulting tilt is different across locations.
Pixels that are occluded by the front focal plane need to be tilted the most, and the amount of tilt gradually reduces when a pixel moves away from the occluding edge, as shown in \figsym~\ref{fig:idea}d.

Despite its simplicity, \conetilt effectively reduces light leakage across focal planes.
Even though point $a$ is transparent, \conetilt ensures that no pixel on the back focal plane emits light toward point $a$, and thereby, we cannot see the far plane when we look at the front object.  
This effectively creates an illusion that the front object blocks light.
In addition, contrast is preserved as no light leaks through the front object. 
Since entire light cones are tilted, \conetilt does not require additional angular resolution and in principle can have the same spatial resolution as a typical multifocal display. 
A side effect of applying the \conetilt operator is that \conetilt eliminates some light rays of the back layer that are not blocked in reality.
This leads to some dark halo around occlusion boundaries.   
We will explain this in more details in \secsym~\ref{sec: properties}.

\section{Design of ConeTilt Displays} \label{sec:design}

In this section, we describe an optical schematic for implementing the \conetilt operator, as well as derive its content generation rules. 

\begin{figure}
\centering
	\includegraphics[width=0.7\linewidth]{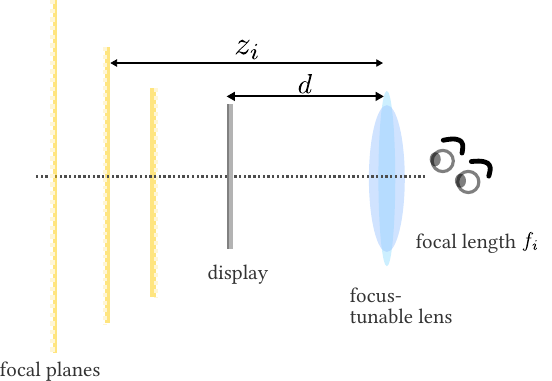}
	\caption{\textbf{Schematic of a standard multifocal display.}  A multifocal display consists of a display placed at a distance $d$ away from a focus-tunable lens, whose focal length is varied to produce multiple focal planes sequentially.}
	\label{fig:multifocal}
\end{figure}

\subsection{Optical Schematic  }
\label{sec: optical schematic}

\subsubsection{Optical Schematic of standard multi-focal displays}
The most popular implementation of a multifocal display uses a focus-tunable lens located at distance $d$ in front of a display panel. When the focus-tunable lens has a power corresponding to focal length $f_i$ it generates a virtual copy of the display, or a focal plane, at distance $z_i$ (see \figref{fig:multifocal}) defined by the thin-lens formula
\begin{equation}
\frac{1}{d} + \frac{1}{-z_i} = \frac{1}{f_i}.
\label{eq: thin lens}
\end{equation}
In other words, when the lens has a focal length  $f_i$, the content is presented to the viewer at the depth $z_i$.
To display multifocal focal planes at different depths, the focus-tunable lens and the display cycle through multiple $\{f_i,z_i\}$ values,  displaying content at each focal plane, within the persistence of vision of the human eye.
The outcome is that the viewer perceives the superposition of content at all focal planes.

\subsubsection{Optical Schematic of the \conetilt display}
\begin{figure}
	\includegraphics[width=\linewidth]{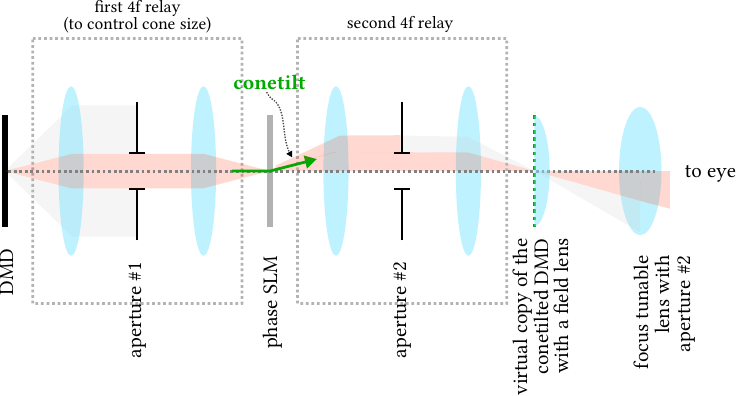}
	\caption{\textbf{Schematic of a \conetilt multifocal display.} We implement the \conetilt operation by optically colocating a phase SLM with a display panel (DMD). This is achieved by mapping the physical display onto the phase SLM using a $1{:}1$ $4f$ relay. The phase SLM implements the \conetilt operator. Subsequently, a second $4f$ relay is used to map the phase SLM onto the image plane of the focus tunable lens.}
	\label{fig:optic-setup}
\end{figure}

The \conetilt operation is implemented by optically attaching a phase SLM to the display panel, a digital micromirror device (DMD) in our prototype.
This optical setup is illustrated in \figref{fig:optic-setup}, which is composed of the DMD, the phase SLM, two one-to-one $4f$ relays, a field lens, and a focus-tunable lens that serves as the main lens of the multifocal display.
The first $4f$ relay optically colocates the DMD and the phase SLM, and the second relay is used to place a field lens which will be discussed below.
We also use the extra spacing introduced by the second relay to place additional calibration cameras (please see the supplemental material for details).
Conceptually, as the phase SLM is optically collocated on the DMD, it serves as a free-form field lens and only controls the direction of the light from the DMD without introducing any magnification that will reduce the spatial resolution of the display.
The aperture of the first $4f$ relay ensures a fixed angular cone arriving the SLM from all DMD pixels. 
Note that we need to crop any tilted light ray whose direction exceeds the angular range of the original light cone.
This is achieved with a second aperture that can be placed at the second $4f$ relay or on the focus tunable lens.

\subsection{Deriving the Direction and Magnitude of the Cone Tilt} 
\label{sec:parameters}
\begin{figure}
	\begin{subfigure}{\linewidth}
	\centering
		\includegraphics[width=0.9\linewidth]{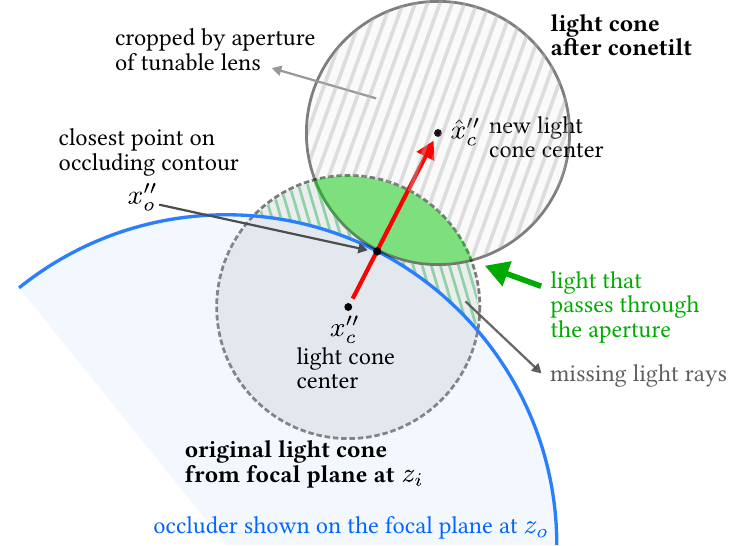}
		\caption{View on the focal plane at depth $z_o$}
		\label{fig: cone shift}
	\end{subfigure}
\\
	\vspace{3mm}
	\begin{subfigure}{\linewidth}
		\includegraphics[width=\linewidth]{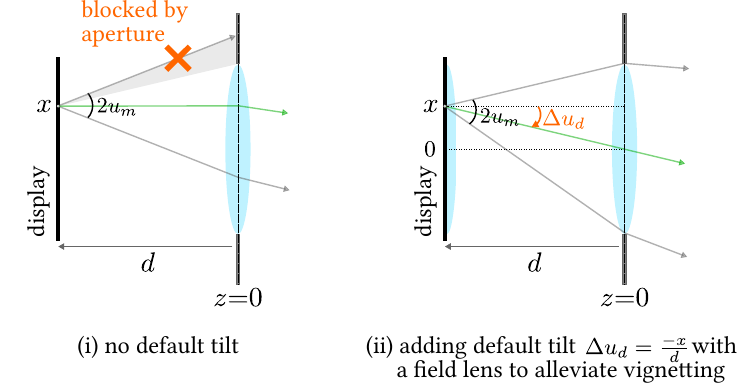}
		\caption{Reducing vignetting with a field lens}
		\label{fig: vignette}
	\end{subfigure}
	\\
	\vspace{3mm}
	\begin{subfigure}{\linewidth}
		\includegraphics[width=\linewidth]{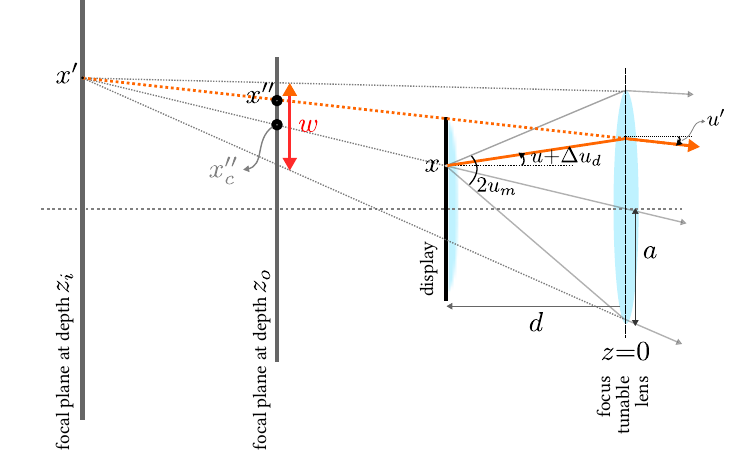}
		\caption{Ray diagram}
		\label{fig: notation}
	\end{subfigure}
	\caption{\textbf{Determining \conetilt parameters.} (a) shows the intersection of the light cone on the focal plane at depth $z_o$ where the occluder (blue region) is placed. The \conetilt operator  shifts the light cone by the smallest amount such that it does not overlap with the occluder. Due to the aperture of the tunable lens, the slashed gray regions on the tilted light cone is cropped, and  only the light in the solid green region is seen by the viewer.  The slashed green regions represent  light that cannot be rendered by the display; this leads to  a dark halo at depth discontinuities (see \secsym~\ref{sec: properties} for details). (b) Adding a field lens on the display plane reduces the effect of vignetting. 
	(c) We provide the ray diagram, along with the key variables of interest. 
	}
	\label{fig:calculation}
\end{figure}

We now describe our strategy for determining the parameters of the tilt, namely its direction and magnitude, at each pixel.
This is illustrated in \figsym~\ref{fig: cone shift}.
Suppose that a light cone is occluded (partially) by a virtual object on a front focal plane.
The goal of \conetilt is to ensure that no light rays in the light cone intersects with the occluder.
Let the center of the light cone on the front focal plane be $x_c''$.  
We first identify the point $x_o''$ on the occluding contour that is closest to $x_c''$.
Then we steer $x_c''$ towards (or away from) $x_o''$ such that the tilted light cone just touches the occluding contour.
As can be seen from \figsym~\ref{fig: cone shift}, using \conetilt enables the display to approximate the occlusion caused by the virtual object. 

\subsubsection{Image Formation in \conetilt Displays.} 

We derive analytical expressions of the position of the light cone, taking into account the effect of the focus-tunable lens and the vignette of the optics.
For simplicity, we assume small-angle (paraxial) scenarios.

\paragraph{Avoiding Vignetting with a Field Lens.}

Consider a multifocal display which is composed of a focus-tunable lens and a display panel, as shown in \figsym~\ref{fig: vignette}(i).
Without any tilt, the light cone from each display pixel travels straight, and part of the cone will be blocked by the aperture of the tunable lens, which causes vignetting.

To avoid vignetting, we place a field lens on the virtual copy of the display to introduce a default tilt so that the entire light cone enters the aperture without being blocked.
This default tilt directs the chief ray of the light cone towards the center of the tunable lens, as shown in \figsym~\ref{fig: vignette}(ii).
It is easy to show that the default tilt, denoted by $\Delta u_d$, reduces to 
\begin{equation}
\Delta u_d = \frac{-x}{d},
\label{eq: u_d}
\end{equation}
and this is achieved by choosing the focal length of the field lens to be equal to $d$.
This tilt can also be implemented by the phase SLM, but we avoided this due to the limited angular range of the SLM.

\paragraph{Ray Tracing.} 

For simplicity, let us first consider a two-dimensional flatland. %
According to \equsym~\eqref{eq: thin lens}, when the focal-length of the tunable lens is $f_i$, the pixel $x$ on the DMD forms a virtual pixel $x'$ on the focal plane at depth $z_i$, where %
\begin{equation}
x' = \frac{z_i}{d} x,
\label{eq:x'}
\end{equation}
as illustrated in \figsym~\ref{fig: notation}.
This means that after the focus-tunable lens, the light ray $(x,u + \Delta u_d )$ will intersect the focal plane at depth $z_i$ on $x'$ with angle $u'$, where $u \in [-u_m, u_m]$ is the direction of the light ray and $\Delta u_d$ is the default tilt. %
Given the focal length of the tunable lens $f_i$ and \equsym~\eqref{eq: u_d}, we can calculate $u'$ by simple ray tracing:
\begin{equation}
u' = \frac{x + (u + \Delta u_d ) d - x'}{z_i} =  - \frac{x}{d} + \frac{d}{z_i} u.
\label{eq:u'}
\end{equation}

We are interested in the intersection of the light cone on a front focal plane at depth $z_{o} < z_i$ where an occluder lies on.  
Let the intersection of the light ray $(x',u')$ on the front focal plane be $x''$.  
By ray tracing, we have 
\begin{equation}
x'' = x' + u' (z_i - z_o) =  \frac{z_o}{d}x + d z_o \left( \frac{1}{z_o} - \frac{1}{z_i} \right) u.
\label{eq: x''}
\end{equation}

\sloppy From \equsym~\eqref{eq: x''}, we can see that $x''$ is an affine function of $u$, which has two implications. 
First, this means that the light cone $\left\{(x,u) \, | \, u{\in} [-u_m,u_m]\right\}$ intersects continuously on the front focal plane, centers at $x_c''$ ($u=0$), and has a diameter of $w$ given as 
\begin{equation}
x_c'' =  \frac{z_o}{d}x, \quad w = 2 d u_m  z_o \left( \frac{1}{z_o} - \frac{1}{z_i} \right).
\label{eq: w}
\end{equation}
This enables us to calculate whether a light cone is occluded by a virtual front object.
Second, when we tilt the light cone emitted by pixel $x$ by $\Delta u$, the center of the light cone $x_c''$ shifts by 
\begin{equation}
\Delta x_c'' = d z_o \left( \frac{1}{z_o} - \frac{1}{z_i} \right) \Delta u.
\label{eq: delta xc}
\end{equation}
This enables us to compute the required tilt to avoid a front occluder.
The expressions  extend gracefully to 3D scenes by indexing  points with $(x, y)$, instead of $x$, and  angles with $(u_x, u_y)$ instead of $u$.

\paragraph{Calculating \conetilt Parameters}
If a light ray from a back focal plane intersects with a point with content on a front focal plane, we need to design a tilt that will move the cone to the periphery of the content in the front focal plane.
Our strategy is illustrated in \figsym~\ref{fig: cone shift}.
In particular, if a pixel $x''$ on the front focal plane is within distance $w$ from $x_c''$,  the light cone from $x'$ is occluded, and we need to tilt the cone. 
To estimate the required amount of the tilt, $\Delta u$, we first identify the point $x_o''$ on the occluding contour that is closest to $x_c''$; the direction of the tilt is along the vector $x_o'' - x_c''$.
The magnitude of the tilt is determined such that the trailing edge of the cone is incident on $x_o''$ and hence, we can identify the point $\widehat{x}_c''$, that represents the center of the light cone after \conetilt.
From \equsym~\eqref{eq: x''}, we can calculate the tilt $\Delta u_t$ by solving:
\begin{equation}
\widehat{x}_c'' = x_c'' + d z_o \left( \frac{1}{z_o} - \frac{1}{z_i} \right)  \Delta u_t.
\end{equation}
We repeat this for all points on each of the focal planes and effectively compute the \conetilt associated with each displayed pixel.
Having derived the desired tilt for each pixel, we now derive the SLM phase function realizing the tilt.

\subsection{Deriving the Phase Function}
\label{sec:derive_phase}

We start by deriving the ideal phase function and account for SLM restrictions later. 
Let the phase function of the phase SLM be $\phi(x)$, where $x \in \R^2$, and the wave number be $k = \frac{2 \pi}{\lambda}$, where $\lambda$ is the wavelength of the emitted light, which is assumed to be monochromatic or narrowband.
When a light ray reaches the phase SLM at $x$ with direction $u \in \R^2$, the phase function delays the wavefront of the light and causes the light ray to change direction. 
Assuming all angles are small, the outgoing direction $u_o$ can be calculated by
\begin{equation}
u_o = u + \frac{1}{k} \grad{\phi(x)}, \mbox{ or  } \Delta u = \frac{1}{k} \grad{\phi(x)}.
\label{eq: dphi_dx}
\end{equation}
Thereby, our goal is to find a phase function that satisfies $\frac{1}{k} \grad{\phi(x)} = \Delta \bu_t(x) $, where $\Delta u_t(x)$ is the desired tilt of the display pixel at $x$.

We find the phase function by solving a Poisson optimization problem. 
Let $\Delta \bu_{t}^x \, {\in}\, \R^{n_x \times \n_y}$ and $\Delta \bu_{t}^y \, {\in} \, \R^{n_x \times \n_y}$ be the vectorized $x,y$ coordinates of the target tilts  of all display pixels, where $n_x$ and $n_y$ are the number of pixels in the $x$ and $y$ direction, respectively.
Let $\bphi \, {\in} \, \R^{(n_x+1) \times (\n_y+1)} $ be the discretized phase function that we try to find.
We solve the following optimization problem
\begin{equation}
\min_{\bphi} \ \|D_x \bphi  -\Delta \bu_{t}^x\|^2 + ||D_y \bphi  - \Delta \bu_{t}^y||^2  + \epsilon ||\bphi ||^2,
\label{eq: opt}
\end{equation}
where $D_x$ and $D_y$ represent taking derivative along $x$ and $y$, respectively, and $\epsilon$ is a small constant used to control the smoothness of the phase function.
\edit{
In all our experiments, we set $\epsilon = 0.001$.
}

\paragraph{Incorporating Phase SLM Constraints}

Due to the discretization, the phase functions that can be displayed on a phase SLM is limited by the Nyquist sampling theorem.
To avoid phase aliasing effect, we can only show phase functions that do not have high-frequency variations.
The maximum phase difference between two neighboring SLM pixels cannot be more than $\pi$, leading to 
\begin{equation}
\left| \dv{\phi(x)}{x}  \right|  \le \frac{\pi}{\delta_x},
\label{eq: dphi_dx constraint}
\end{equation}
where $\delta_x$ is the pixel pitch of the SLM pixels along the $x$ direction.  
The same constraint applies to the $y$ direction.
The constraint~\eqref{eq: dphi_dx constraint} limits the maximum angle that we can shift the light cones using the phase SLM. 

From \figsym~\ref{fig: cone shift} we can see that given the radius of a light cone $u_m$, the maximum amount of tilt that we will need is $2 u_m$ (when the entire cone goes out of the aperture).
Therefore, \equsym~\eqref{eq: dphi_dx constraint} sets an upper bound of the radius of the light cone:
\begin{equation}
u_m \le \frac{\pi}{2 k \delta_x}.
\label{eq: um upper bound}
\end{equation}
By controlling the first aperture in \figsym~\ref{fig:optic-setup} we can bound 
the light cone to satisfy \equsym~\eqref{eq: um upper bound}.
Our phase SLM has a pixel pitch $\delta_x = 6.4$ \um; when $\lambda = 520$ \nm, the radius $u_m$ is upper-bounded by $1.2$ degrees.
In addition to constraining the size of the aperture used, the limited tilting power of the phase SLM also constrains the eye box of the display, as we will discuss this next.

\begin{figure}
	\includegraphics[width=\linewidth]{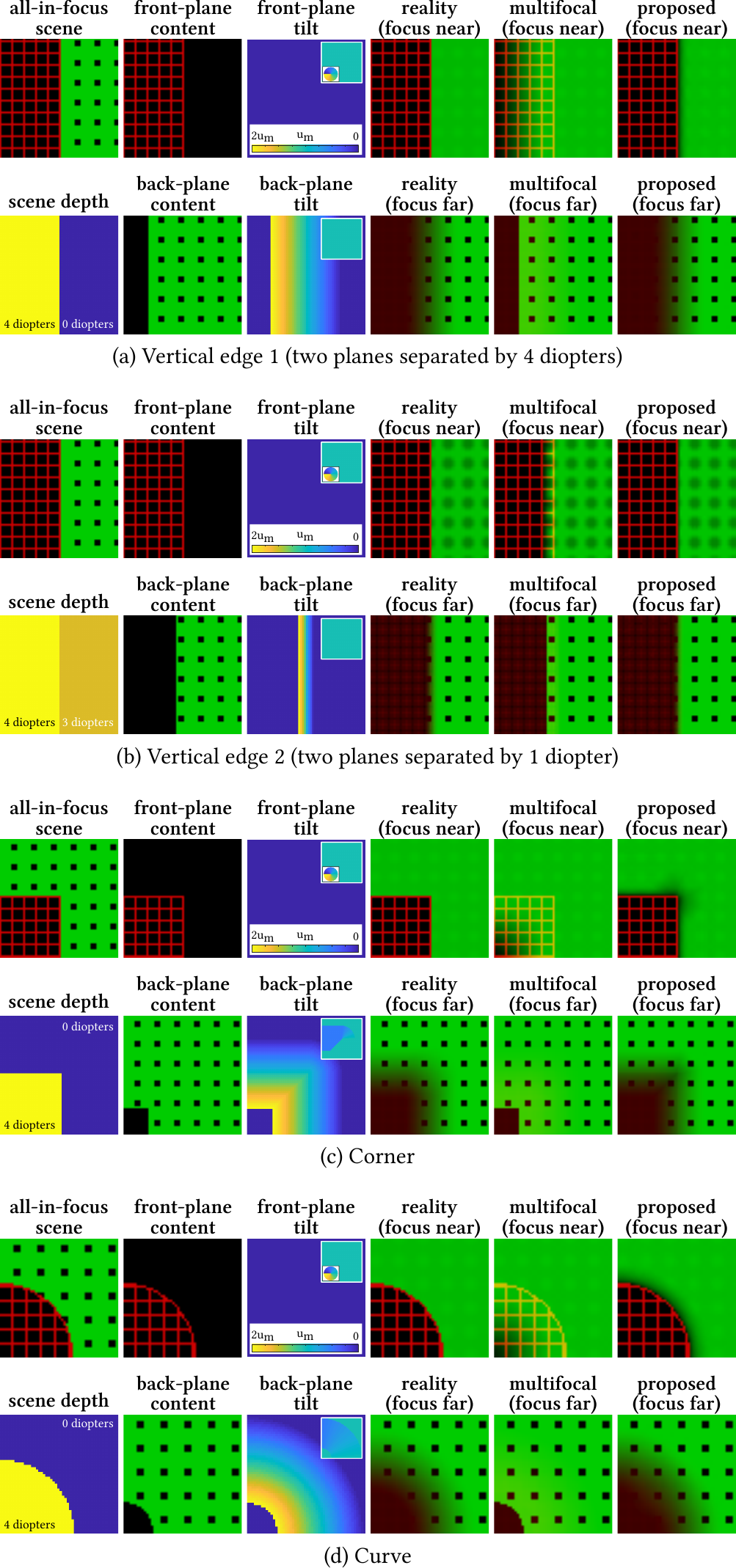}
	\vspace{-5mm}
	\caption{\textbf{Examples.} This figure shows four example scenes, the content shown on the focal planes, the tilt vectors shown with the front and the back plane, the rendered scenes in reality, and the rendered results on a typical multifocal display and our \conetilt display with the same parameters as our prototype.  Note that we plot the tilting vectors in length and direction (insets).  We can clearly see the loss of occlusion cue and light leakage in the typical multifocal display.  The proposed \conetilt display successfully prevents light leakage and creates occlusion at the price of modest dark halo around the occluding contour.}
	\label{fig:examples}
\end{figure}

\paragraph{Examples.}  
 \figsym~\ref{fig:examples} demonstrates some simple  scenes composed of two planes at different depths.  
Given a scene, we find the minimum tilt for each pixel on the back plane to avoid front objects and construct a phase pattern for realizing the tilt. 
As can be seen from renderings of a camera focused on the front and back focal planes, \conetilt effectively avoids the light leakage from the back plane.  

\subsection{Properties of the \conetilt Display}
\label{sec: properties}

\paragraph{Dark Halo Near Occluding Edges.}
Let us revisit the illustrations shown in \figsym~\ref{fig:calculation}a.
In a real scene, the light cone emitted by the background point will be cutoff by the occluder such that only the light in the crescentic region can pass.
In contrast, a \conetilt display --- which works by tilting a small light cone --- can only render the light in the green region, and as a result, some light rays are missing in the virtual scene.
The main effect of missing some light rays, as illustrated in \figsym~\ref{fig:examples}, is that the defocused objects near the occluding boundaries are dimmer compared to the reality.
Note that by reducing the amount of tilt, we can decrease the dark halo. 
This provides an interesting trade-off between the light leakage and dark halo and is left as a future work.

\paragraph{Field-of-View}
When the eye is close to the tunable lens, the field-of-view of a  multifocal display depends on the size of the display panel and the distance $d$.
Our prototype, due to its use of a field lens to avoid vignetting, is capable of displaying content on the entire display panel without being constrained by the phase SLM.
Hence,  its field-of-view is the same as a typical multifocal display of the same design parameters.

\paragraph{Eyebox.} %

Most multifocal displays have small eye boxes, due to the lack of occlusion cues (which causes virtual objects to overlap when the eye shifts). 
As a consequence, even though in principle multifocal displays do not require gaze tracking to provide accommodation cues, most implementations use gaze trackers to re-render the scene as the location of the eye changes~\cite{mercier2017multifocal}. 

In a \conetilt display, eyes can move freely inside the aperture of the tunable lens without causing overlapping contents.  
This extends the effective eyebox to the entire aperture without the help of a gaze tracker or re-rendering.
In our prototype, the aperture size is only limited by the maximal tilt angle of the phase SLM and is equal to $u_m d  = 2.4$ \mm in diameter.   
As stated earlier, the size of the eyebox is primarily determined by the pitch of the phase SLM and using a device with smaller pitch will enhance the size of the eyebox.

\paragraph{Contrast}
With the ability to prevent light leakage, \conetilt displays preserve the contrast of focal planes.  
\figsym~\ref{fig:idea}(b,d) compare the contrast when we display the same content on the focal planes on a typical multifocal display and on a \conetilt display.
As can be seen in the third row,  \conetilt not only  reduces the contribution from the back focal plane to the front focal plane, but also makes the transition  sharper.
Similar trends can be observed in \figsym~\ref{fig:examples}.

\subsection{Relationship to Optimization-based Filtering}
\begin{figure}
	\includegraphics[width=\linewidth]{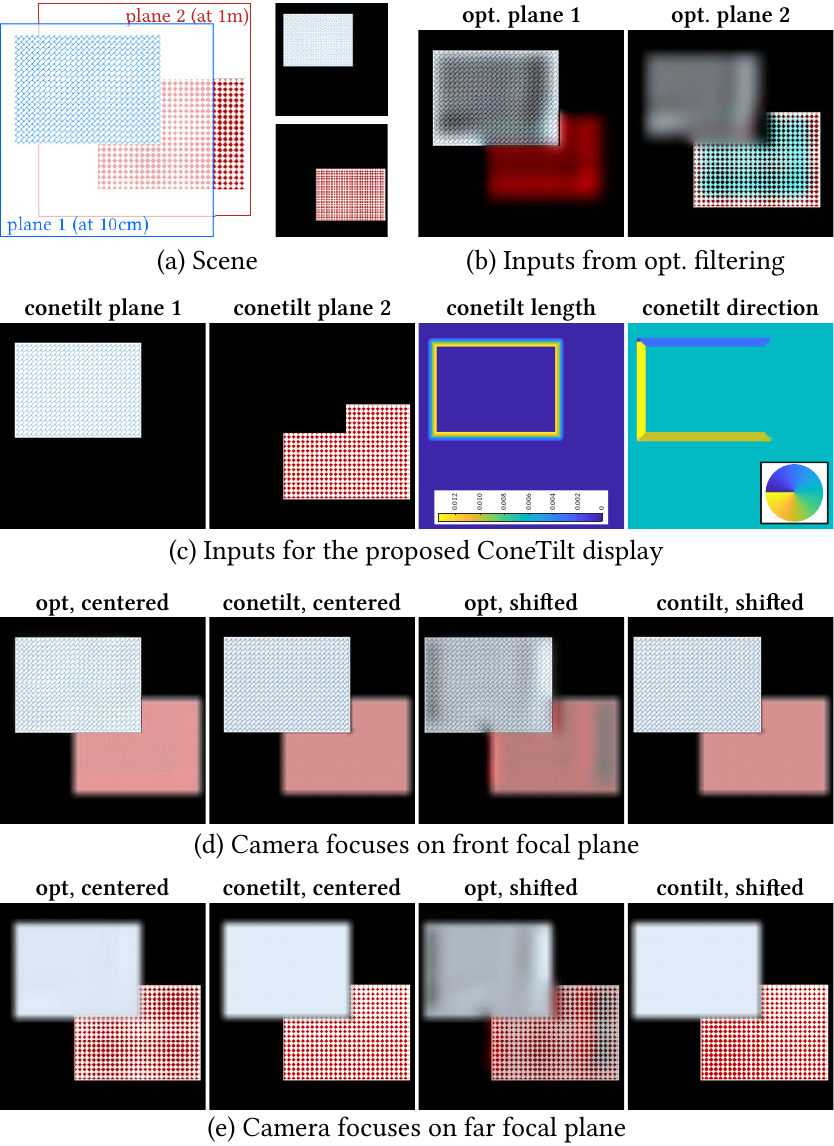}
	\caption{\textbf{Comparison between optimization filtering and \conetilt.} (a) A scene containing two planes at 10 \cm and 1 m. (b) content created by optimization-based filtering. (c) input content for the proposed \conetilt display. We focus the camera (d) on the front plane and (e) on the back plane at both the center position and a slight shift to the right. The shift causes the back plane to move by 1 pixel and the front plane to move by 10 pixels in the same direction. \edit{This corresponds to a 2 \mm shift on a VR headset with a horizontal resolution of 1080 pixels and a $100^\circ$ field-of-view.} The view from the optimized display degrades  as viewpoint shifts, while  \conetilt faithfully reproduce the new viewpoint.  }
	\label{fig:opt comparison}
\end{figure}

\edit{\conetilt displays can also be interpreted as a hardware counterpart to optimization-based content generation~\cite{akeley2004stereo,narain2015optimal,mercier2017multifocal,choi2019optimal,xiao2018deepfocus} for handling  the transparency of focal planes in a multifocal display.}
\citeN{narain2015optimal} show that leakage of defocus blur at depth discontinuities can be alleviated by optimizing the content  shown at the different focal layers.
However, since  this approach results in a single object being rendered on multiple focal planes, small motion of the eye can lead to inconsistent motion parallax and occlusion cues unless the content is regenerated, using an eye and head tracking system \cite{mercier2017multifocal}.
\sloppy {
We test the effectiveness of \conetilt displays and the approach of \citeN{narain2015optimal} in \figsym~\ref{fig:opt comparison}.
Even though the optimization-based filtering successfully reproduces the scene when the eye is centered, the quality of the results deteriorates with a slight viewpoint change.}

\section{Prototype and Experimental Setup}
\label{sec: prototype}

\begin{figure}
	\includegraphics[width=\linewidth]{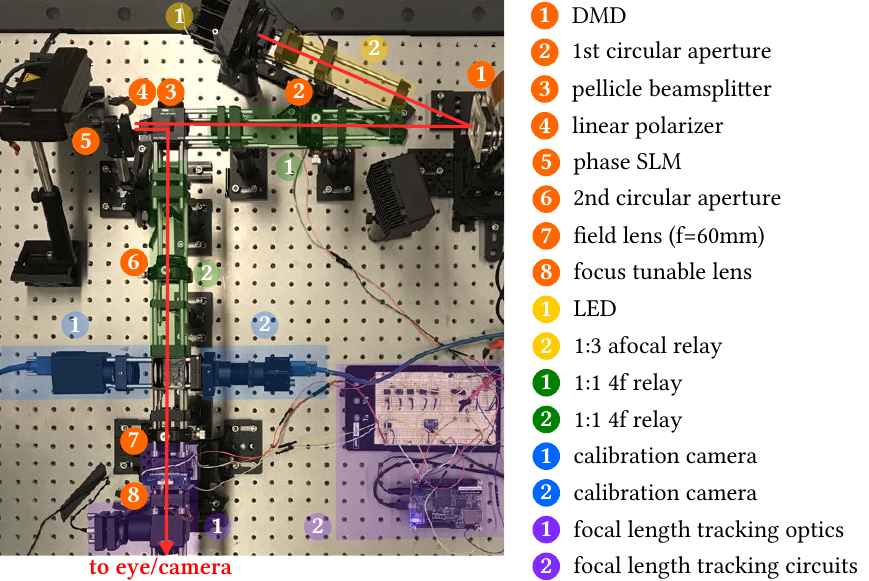}
	\caption{\textbf{Lab prototype.} We implement the schematic of \figref{fig:optic-setup} using off-the-shelf components.  The list of components is provided in the supplemental material.
}
	\label{fig:prototype}
\end{figure}

We follow the schematic shown in \figsym~\ref{fig:optic-setup} and build a prototype \conetilt display shown in \figsym~\ref{fig:prototype}.
We use a green LED whose spectrum centers at $520$ \nm as our light source, and we calibrate the phase SLM to operate at this wavelength.
Our prototype implements a light cone of $1.2$ degrees in radius, a field-of-view of $6.8$ degrees in diameter, and an eye box of $2.4$ \mm in diameter.
Note that since our prototype uses a physical field lens to implement the default tilt, the field of view is the same as a multifocal display of the same configuration.  
The small field-of-view is due to the simplicity of our implementation and can be increased by moving the tunable lens closer to the phase SLM, \ie, reducing $d$, which is currently $58$ \mm.

\edit{
To control the focus tunable lens, we follow the implementation of \citeN{chang2018towards} and build a focal-length tracking system.  
Our prototype is capable of displaying up to 40 focal planes, uniformly separated (in diopter) from 0 to 4 diopters. 
We discuss the implementation details in the supplemental material.
}

\subsection{Display Inputs and Capturing Process} \label{sec: display}

\paragraph{Inputs.}

Given a 3D scene, we first discretize the scene according to the depth of the focal planes (in diopters) and assign each point in the scene to its nearest focal plane, as our system has a high depth sampling rate (10 focal planes per diopter), depth discretization introduces minimal errors visually.
Given the size of the light cone, we remove all pixels that are  completely occluded (not seen along each ray in the cone).
We then follow the algorithm described in \secsym~\ref{sec:parameters} to compute the tilt for each pixel and the phase function to show with each focal plane.

To evaluate the effectiveness of \conetilt, we do not apply any depth filtering (\eg, linear or optimization-based filtering) to the content.
\edit{
	Nevertheless, for most of the scenes, including Figs.~\ref{fig:teaser}, \ref{figure: sunny}, \ref{fig:psf_aperture}, \ref{figure: chess}, and \ref{fig:opt comparison}, the focal planes of the display match the depth planes of the content such that a virtual object lies entirely on a focal plane; for this specific scenario, linear depth filtering~\cite{akeley2004stereo} will have no effect on the input images.
}

\begin{figure}[t]
	\centering
	\includegraphics[width=\linewidth]{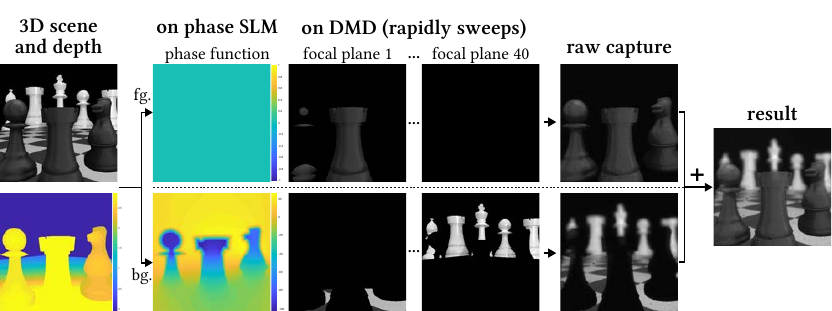}
	\vspace{-6mm}
	\caption{\edit{\textbf{Display and capturing process.} Given a 3D scene and its depth
map, we first decompose the scene into foreground and background,  to be displayed using two different phase functions. Foreground/background content are associated to the corresponding depth planes.  The phase function is fixed during each sweep. The final result is the sum of the two captured images. Note that the phase function and the content on the focal plane are entirely independent to the camera.}}
	\label{fig: capture procedure}
\end{figure}

\edit{
\paragraph{Display process.} 
In a straightforward implementation of a multifocal display, one sweeps through the planes, displaying the intensity content of each plane with the DMD and the phase content with the SLM. 
However, while the refresh rate of the DMD and focus tunable lens are high, the refresh rate of the SLM is limited. 
To bypass this limitation, we noticed that most scenes can be displayed using only two phase patterns: one pattern displaying the pixels on the foreground (i.e., unoccluded) pixels and one pattern for the background that includes content that is occluded by the foreground for at least one ray via the aperture, namely all the pixels which should be tilted.  
A single frame of the VR content is, therefore, displayed with two sweeps of the focal  tunable lens; in the first sweep, the DMD shows the foreground and the phase SLM pattern is set to zero, and in the second sweep, the background is shown on the DMD with the \conetilt phase pattern on the SLM.
This implies that when the focus tunable lens and DMD display the content of a particular focal plane, the phase tilt associated with all other planes is on as well. However, as no content is shown by the DMD at that part of the frame, the tilt of other focal planes does not contribute to the final image.
Figure \ref{fig: capture procedure} shows the images displayed on the DMD and SLM during these two cycles, and the images captured by the camera observing them.}

\paragraph{Capture process.}
We use a FLIR Grasshopper grayscale camera with a Nikkor 35 mm prime lens to capture the photos.
We use $f/22$ so that the aperture of the lens lies entirely within the eyebox of our prototype. 
The camera is put on a linear translation stage in front of the tunable lens to mimic the eye movement.
We use a $1{:}1 \ 4f$ relay to map the camera to the aperture of the tunable lens.  
This provides ample space for mounting the translation stage and eliminates the magnification due to  the unnecessary distance between the camera and the tunable lens. 
To simplify the synchronization between the DMD and the phase SLM, we capture the foreground and background content \edit{\textit{separately} and sum the two results in post processing, as shown in \figsym~\ref{fig: capture procedure}}.
\newedit{
To capture each of the foreground and background, we capture and average 10 images, each has a different global phase offset ranging from $0$ to $\pi$ (please see \secsym~\ref{sec: slm limitation} for detailed discussion).
We use exposure time equal to $730$ \ms, and the overall capturing process for one grayscale result takes about 20 seconds.
}
Since our prototype is grayscale, to show RGB contents, we display and capture each color channel \edit{\textit{separately}} and assemble a three-channel image computationally.
\edit{
Hence, the results shown in the paper are produced artificially to mimic a field sequential display.
}
Note that during the capturing process, the camera is entirely independent of the display, i.e., we do not re-render the scene based on the camera configuration.

\section{Results} \label{sec:results}

In the following, we show the results  of the \conetilt  display on various scenes designed to highlight the important features of the proposed method. We encourage the reader to check supplemental videos demonstrating translation of viewing position as well as changes of the focal plane.

\subsection{Controlling  Light Cones with \conetilt}

\begin{figure}
	\includegraphics[width=\linewidth]{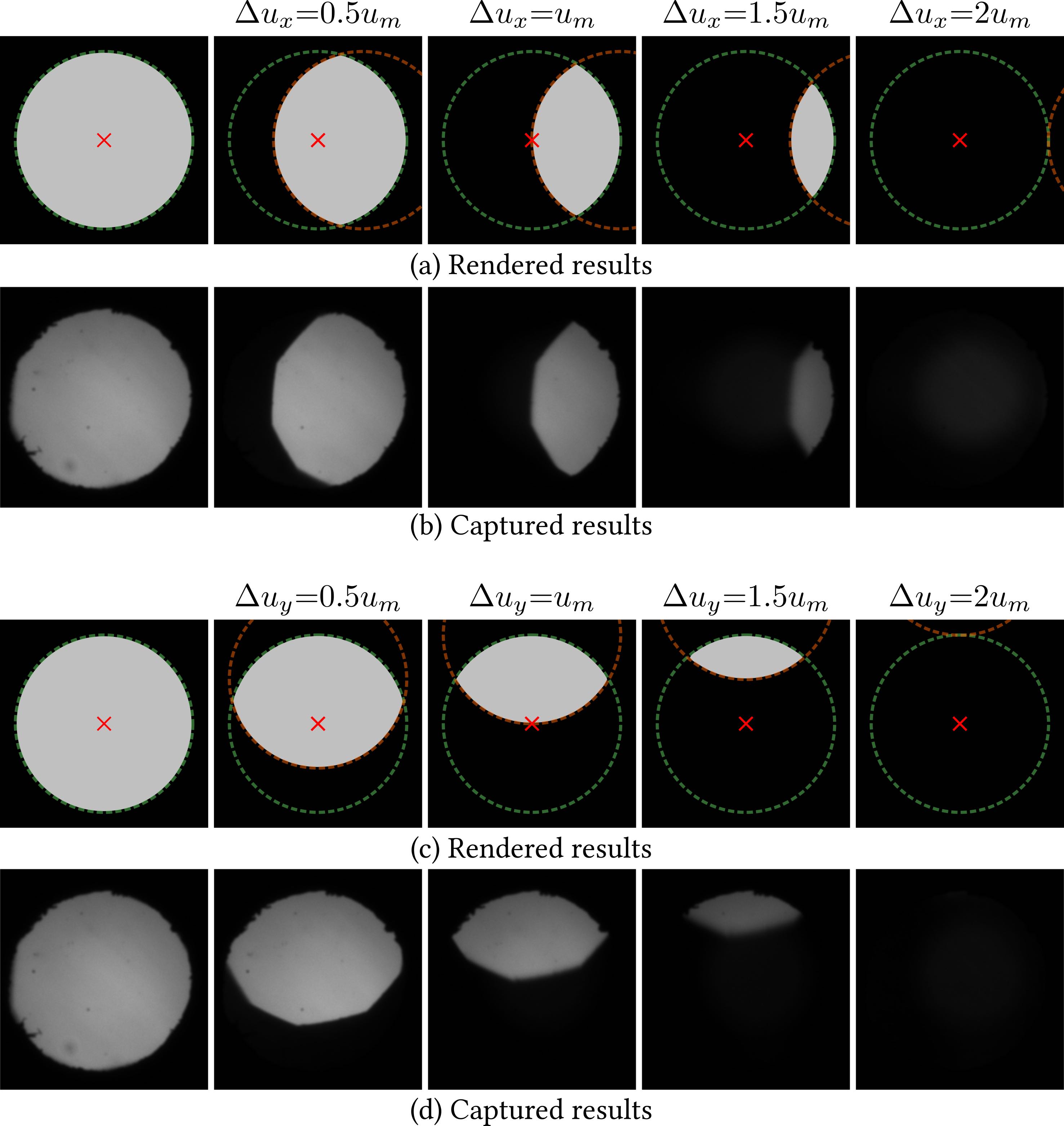}
	\caption{\textbf{Tilted light cones.}  We focus the camera on the aperture of the tunable lens and show an all-one image on the DMD with different global tilting configurations, demonstrating the tilted cones.  }
	\label{fig:psf_aperture}
\end{figure}

We verify the ability of \conetilt to tilt light.
\figsym~\ref{fig:psf_aperture} shows the light entering the tunable lens under different configurations of tilts. 
We show a full white image on the DMD and tilt every pixel in the same direction. 
The results are captured by focusing a camera on the aperture of the tunable lens  demonstrating how the light cone is tilting (for details, see the supplemental material).

\begin{figure}
	\includegraphics[width=\linewidth]{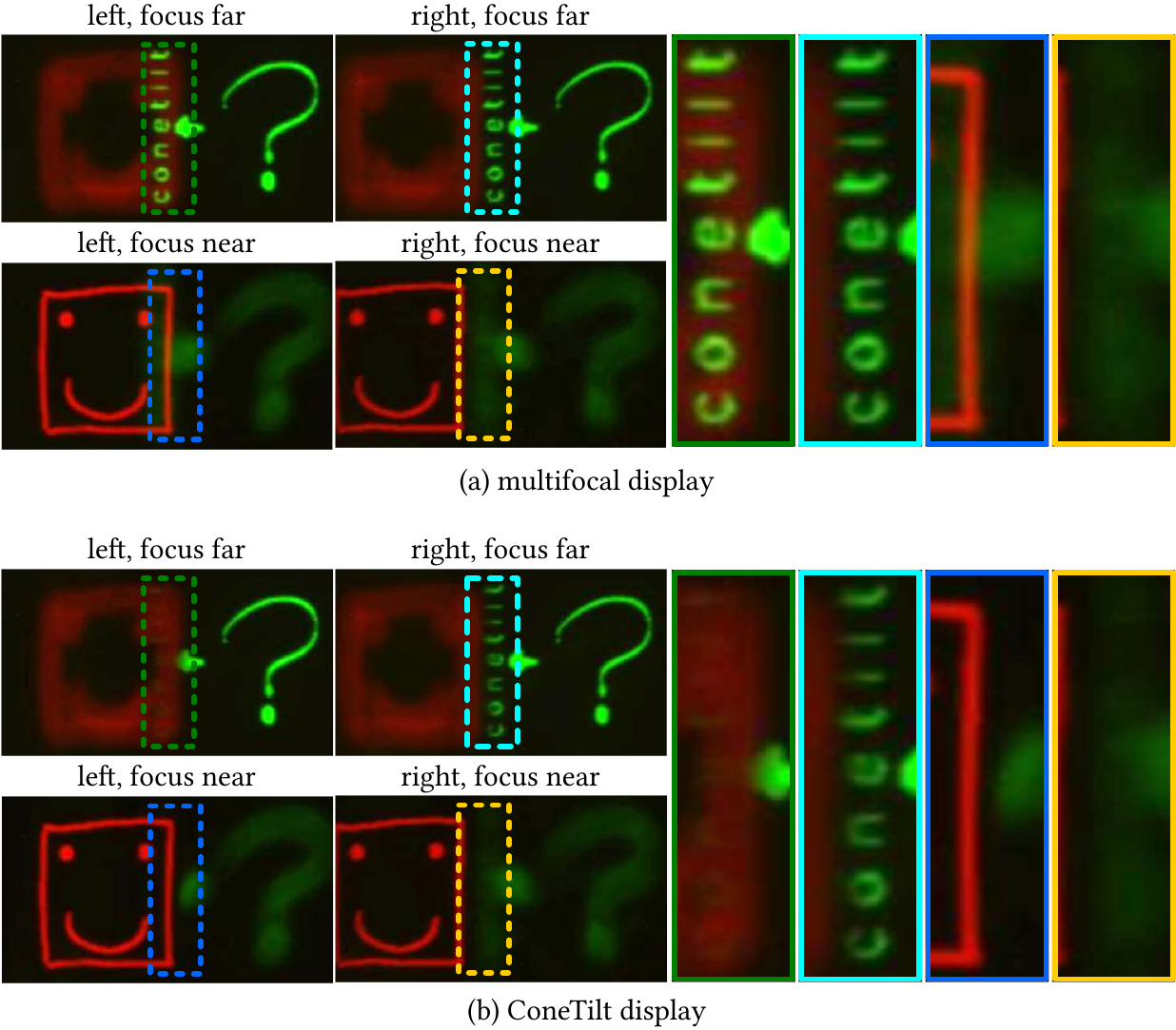}
	\caption{\textbf{Creating occlusion cue.} The figure shows the captured photo of the scene shown in \figsym~\ref{fig:teaser} when the camera is translated \edit{($-0.5$ \mm and $+0.5$ \mm from the optical axis, respectively)}.   The front smiley face is opaque and should occlude the text and part of the arrow when the camera is at the left position.  We show insets of interesting regions to highlight artifacts in traditional multifocal displays.  }
	\label{fig:square head}
\end{figure}

\subsection{Hiding Content Behind  Occluders}
We demonstrate the capability to hide content behind an occluder and reveal it when the camera/eye shifts --- all without re-rendering the scene.  
The scene in  \figsym~\ref{fig:teaser}, \ref{fig:square head} contains an opaque smiley face in the front and a question mark and the text ``conetilt'' in the back.  
We shift the camera with a translation stage from left to right; when the camera is at the left position the text should be occluded by the smiley head, and the text should be revealed when the camera shifts to the right. 
As can be seen from the results, the smiley face rendered by the typical multifocal display fails to occlude the text and even makes the text brighter due to the additive nature of the front and the back focal planes.  
In comparison, the text is occluded and revealed when \conetilt is applied.  
The lower intensity of the text is as expected, since most of the light rays from the text are occluded by the smiley face, as happens in reality.
The results demonstrate the ability of the proposed display to support small shifts of the pupil without the help of a gaze tracker or any additional rendering.  

\subsection{Generic Occluding Contours}

\begin{figure}
	\centering
	\includegraphics[width=\linewidth]{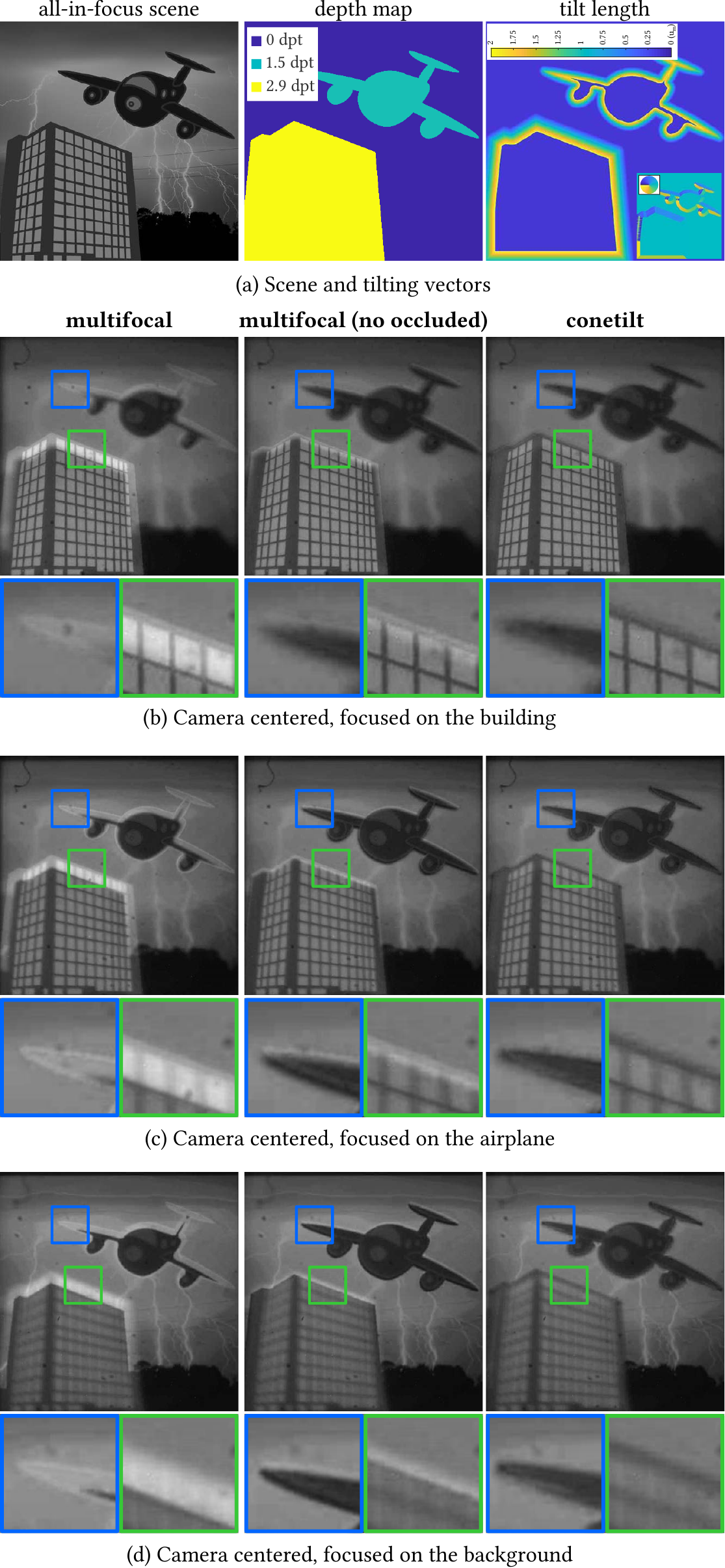}
	\caption{\textbf{Lightning.} The figure shows the captured images of the scene shown on the top left. The tilting vectors are shown on the bottom left with the direction of the tilting vectors shown in the inset. ``no occluded'' means that we remove the directly occluded regions in the background.}
	\label{fig:lightning}
\end{figure}

\begin{figure}
        \centering
        \includegraphics[width=\linewidth]{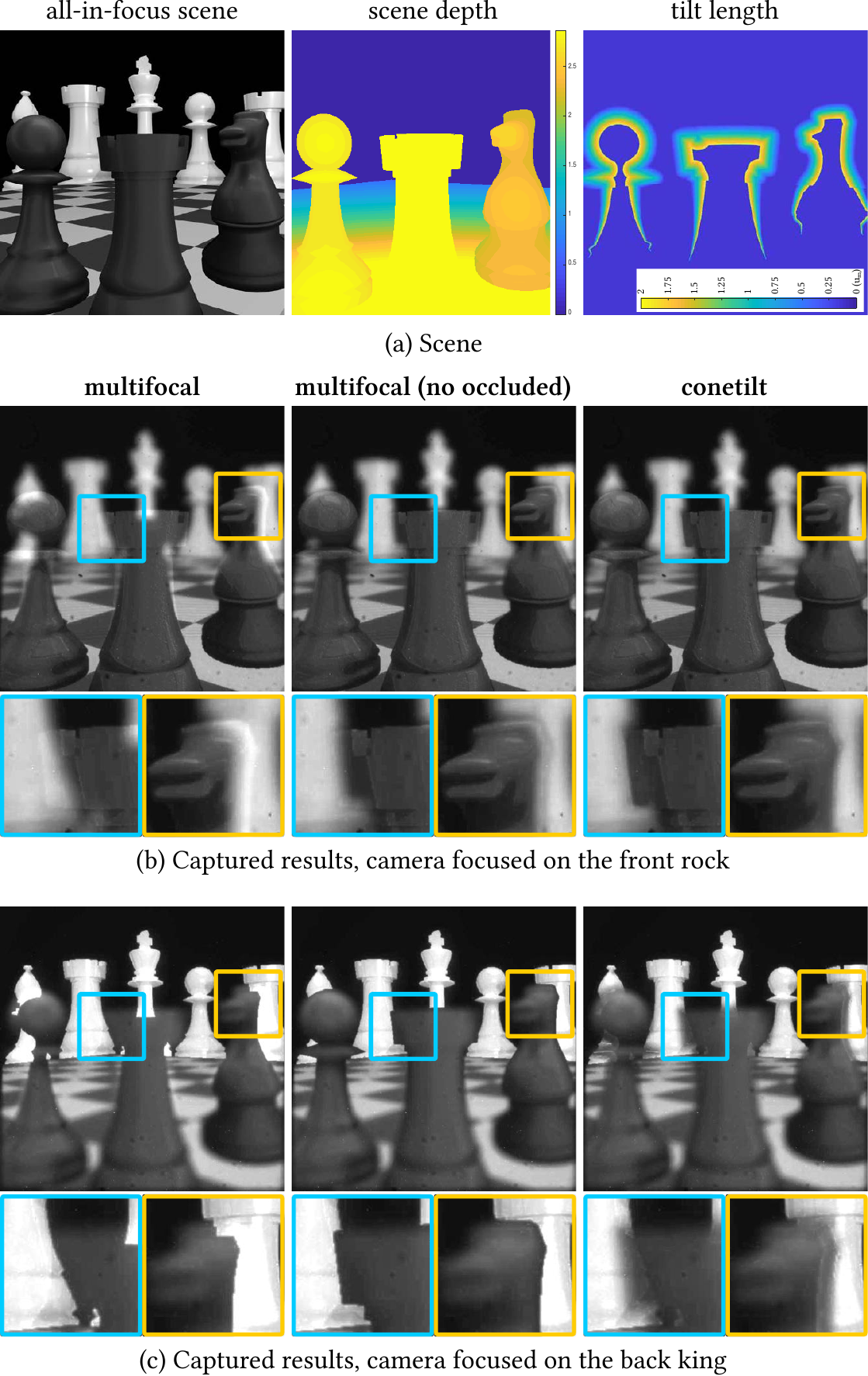}
        \caption{\textbf{Chessboard.} (a) The scene is composed of a chessboard and several chess pieces. All black chess pieces are in the front, and the white ones are in the back. Each pixel in the scene is assigned to one of 40 focal planes according to their depth. (b) shows the captured photos by a centered camera focusing on the front rock piece. On the multifocal displays, with or without showing the directly occluded content, the light leakage from the background etches into the front chess pieces and makes them look smaller. In comparison, the \conetilt display prevents the light leakage and preserves the shape of the front pieces. (c) shows the captured photos when the camera focuses in the back on the king piece.  On the multifocal display, since the front focal plane is transparent, we can directly see the regions in the background that should be occluded.  When we crop out the directly-occluded regions on the back chess pieces, the cropping edges get in focus and become unnaturally sharp. This creates a false illusion that the camera is focusing on the black piece in the front. With \conetilt, the front pieces appear defocused and are able to occlude the white chess pieces in the back.
		}
        \label{figure: chess}
\end{figure}

\begin{figure}
        \centering
        \includegraphics[width=\linewidth]{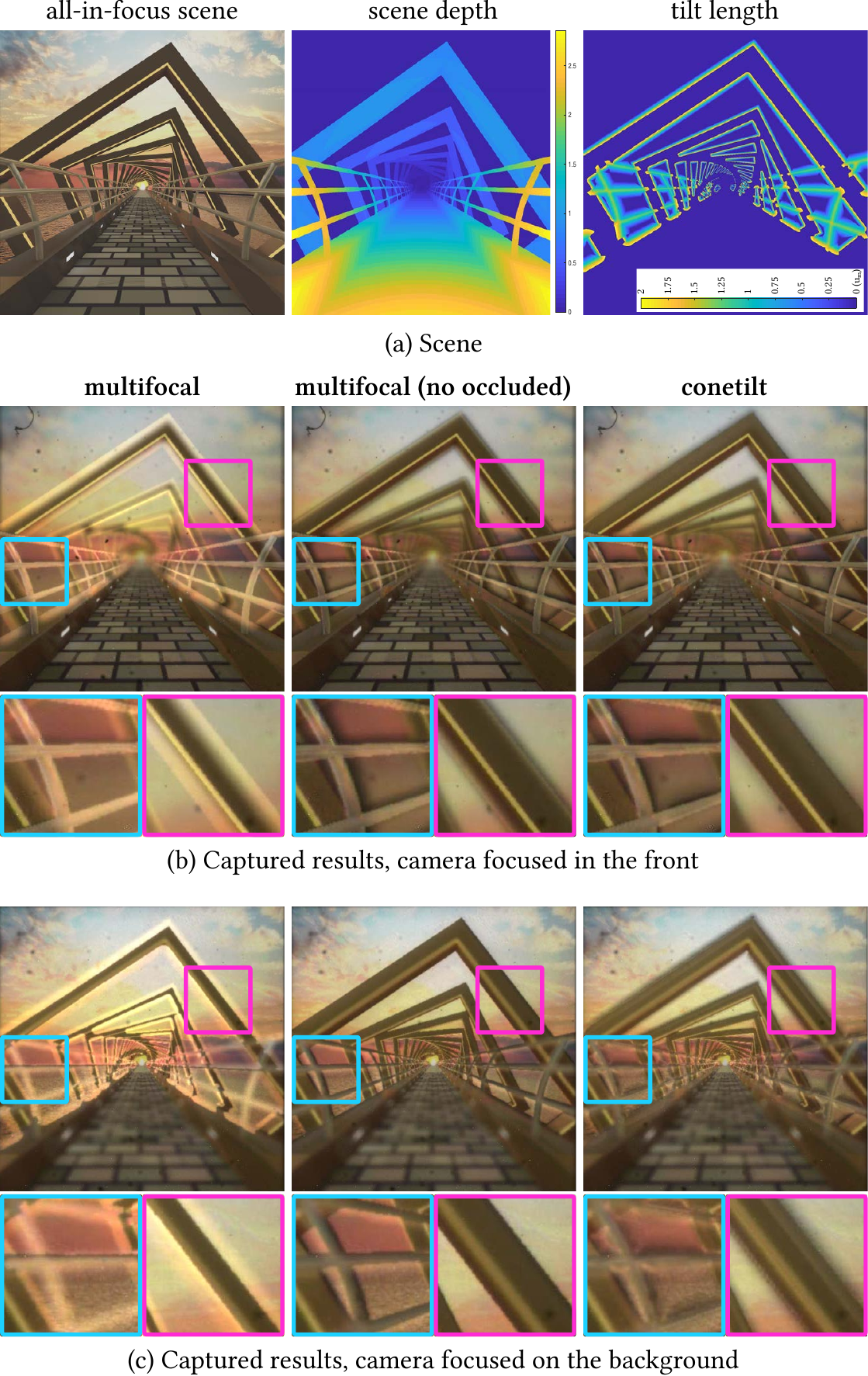}
        \caption{\edit{\textbf{Bridge.} (a) A  scene composed of a bridge with railings, that stretch from the camera to infinity, scaffolding over the bridge at a few depth planes and a sky background at infinity. As with \figsym~\ref{figure: chess}, we discretize the scene into 40 focal planes uniformly in diopters from 0 to 4 diopters, and assign each pixel to its closest focal plane in depth. The  length of the cone tilt is also shown. (b, c) Captured images with multifocal, with  and without directly-occluded regions, and the \conetilt displays. The insets in (b), where the camera is focused in front show significantly reduced light leakage in the \conetilt result, as compared to the multifocal displays. Yet, the close spacing of foreground occluders and the dark halo lead to infeasible cone tilts for some regions, that we highlight in the cyan inset in (c). Such artifacts can likely be avoided with operations more complex than simple tilts.}}
        \label{figure: bridge}
\end{figure}

\begin{figure}
\centering
\includegraphics[width=\linewidth]{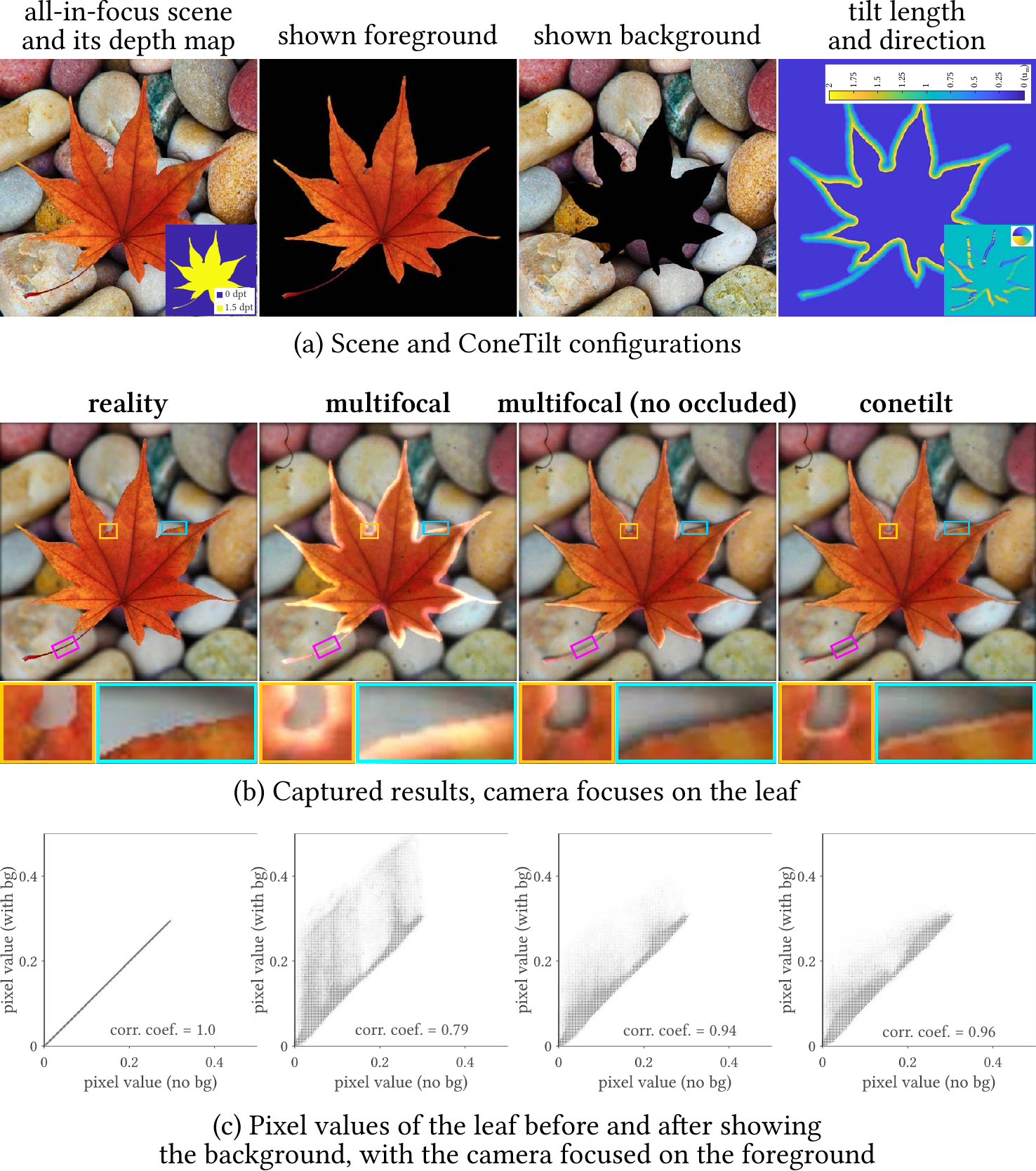}
\caption{\textbf{Leaf.} (a) We present a scene with two focal planes along with the content displayed on the two planes as well as the \conetilt displayed on the back plane. (b) Photographs obtained with different display configurations with the camera focused on the front plane. (c) A scatter plot of intensities observed on foreground pixels with and without the background.  The $x$-axis represents the pixel values when the background is not shown, and the $y$-axis represents the pixel intensities after showing the background. In an ideal display, we expect all point to lie on the $x =y$ line, as it does in the rendered reality.}
\label{figure: leaf}
\end{figure}

We show captured results on scenes with more complicated occluding contours in \figsym~\ref{figure: sunny} and  \figsym~\ref{fig:lightning}-\ref{figure: leaf}.
In \figsym~\ref{fig:lightning} we also compare the captured results against rendering of what one would expect to see in reality. 
From the results, we can make the following observations. 

\paragraph{Reduced Leakage.}
All results consistently demonstrate that the \conetilt display effectively reduces light leaking from the background onto foreground occluders. 
Please see the boundaries of the building in \figsym~\ref{fig:lightning}a, the top of the rock in  \figsym~\ref{figure: chess}b,  the boundary of the leaf in  \figsym~\ref{figure: leaf}b, and the scaffolding structures in \figsym~\ref{figure: bridge}b.

While removing the directly occluded regions in the background helps reduce the light leakage in multifocal displays, it only works for a certain viewing position and angle.
As can be easily seen from the supplemented videos, when the camera shifts left and right \edit{between $\pm 0.5$ \mm}, multifocal displays without directly-occluded content still suffer from light leakage.
In addition, removing directly occluded content also worsens the dark halo, as demonstrated in the results and in \figsym~\ref{fig:idea}c.

\paragraph{Improved Contrast.}
To quantitatively characterize the effect of \conetilt on the contrast of the foreground, in \figsym~\ref{figure: leaf} we capture each of the display options twice, once in its standard mode and again when showing a black image at the background plane. 

In \figsym~\ref{figure: leaf}c we display a scatter plot where the horizontal position of a point corresponds to a grayscale intensity at a foreground pixel of a background-free image, and the vertical position is the intensity of the same pixel when background is displayed.
In reality, since the leaf is opaque and is in focus, showing the background should not affect its pixel values, so we would expect the scatter plot to be a diagonal line ($x=y$).
In practice when the background is shown on a multifocal display, light leakage increases the brightness near the depth discontinuities, resulting in pixels with values above the diagonal line and reducing the correlation coefficient.
In comparison, the histogram produced by the \conetilt display is much closer to the diagonal line and has a higher correlation coefficient.

\paragraph{Defocus Cues.}
The captured results also demonstrate another advantage of \conetilt displays over typical multifocal displays.
When a multifocal display attempts to reduce light leakages by removing directly occluded content on the background, it deteriorates the defocus cue of the occluder when the camera focuses on the background.
As can be seen from \figsym~\ref{figure: chess}c, the defocused foregrounds of the multifocal display (no occluded) look unnaturally sharp even though in reality they should be blurred due to defocus.
In comparison, the \conetilt display successfully renders blurred foregrounds, which is often important for improving the immersion of VR displays~\cite{zannoli2016blur}.

\edit{\paragraph{Dense depth variations.} \figsymp~\ref{figure: chess} and \ref{figure: bridge} show two scenes with a continuous depth variation, which we display with a dense scan of 40 different focal planes, sampled uniformly in diopter from 0 to 4 diopters. Each pixel is assigned to the focal plane with closest depth.}

\edit{
\paragraph{Quantitative performance.} We quantitatively characterize the  performance benefits  of the \conetilt displays. For the dinosaur scene in \figsym~\ref{figure: sunny}, over ten rendered images with different viewpoints, we observed an average PSNR of 23.5 dB and SSIM score of 0.967 for a traditional multifocal display, when compared to ground truth renderings. \conetilt renderings achieved an average PSNR of 31.2 dB and SSIM score of 0.986. The small quantitative difference can be attributed to the depth boundaries being sparse.}

\section{Discussions}
\label{sec: limitation}
We discuss some of the features of  \conetilt displays, including key limitations  and potential ways to mitigate them, as well as approaches to miniaturize our prototype and obtain a form factor suitable for VR glasses.

\begin{figure}[!ttt]
	\centering
	\begin{subfigure}{0.47\linewidth}
		\centering
		\includegraphics[width=0.9\linewidth]{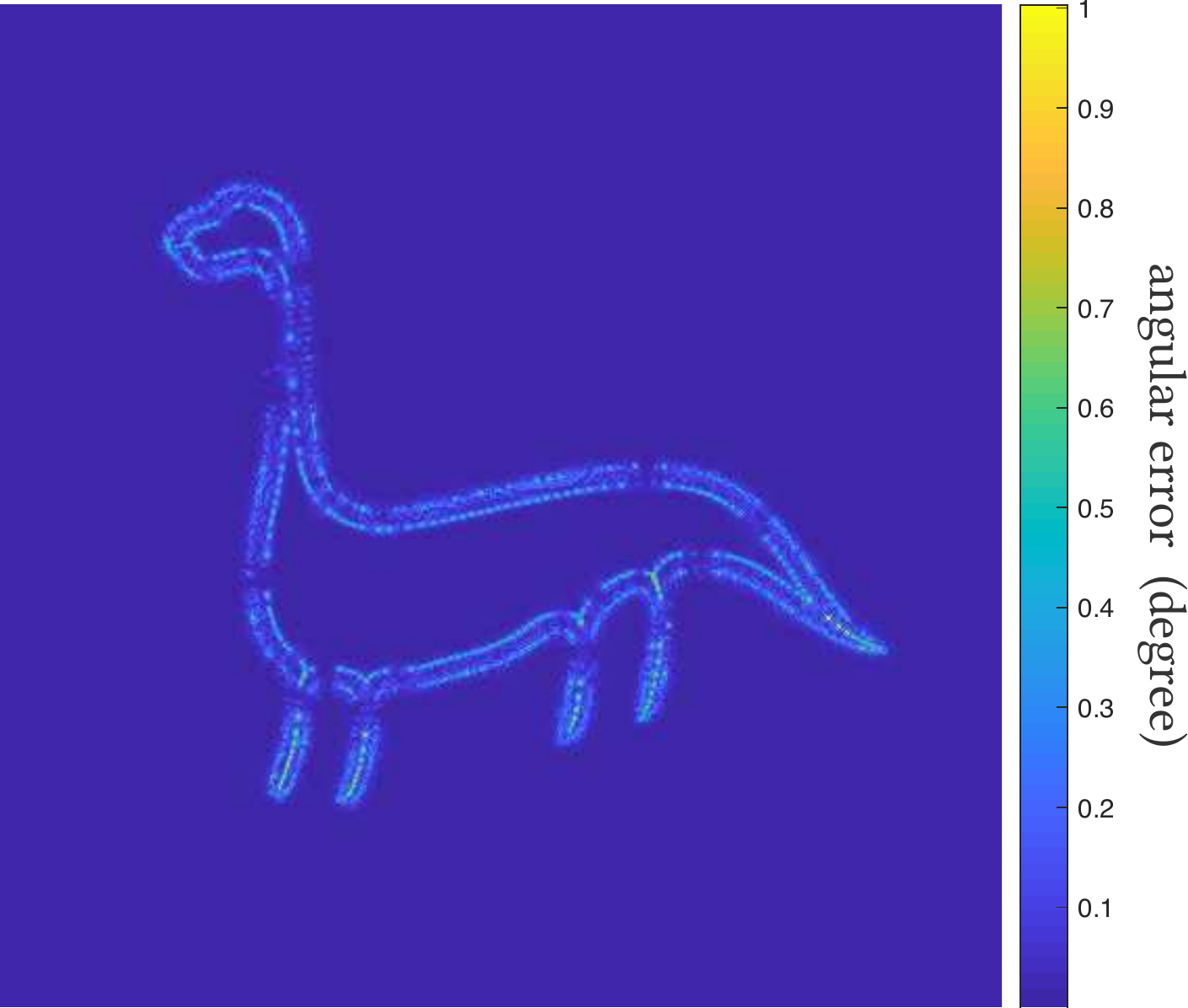}
		\caption{Scene: \figsym~\ref{figure: sunny}}
	\end{subfigure}
	\begin{subfigure}{0.47\linewidth}
		\centering
		\includegraphics[width=0.9\linewidth]{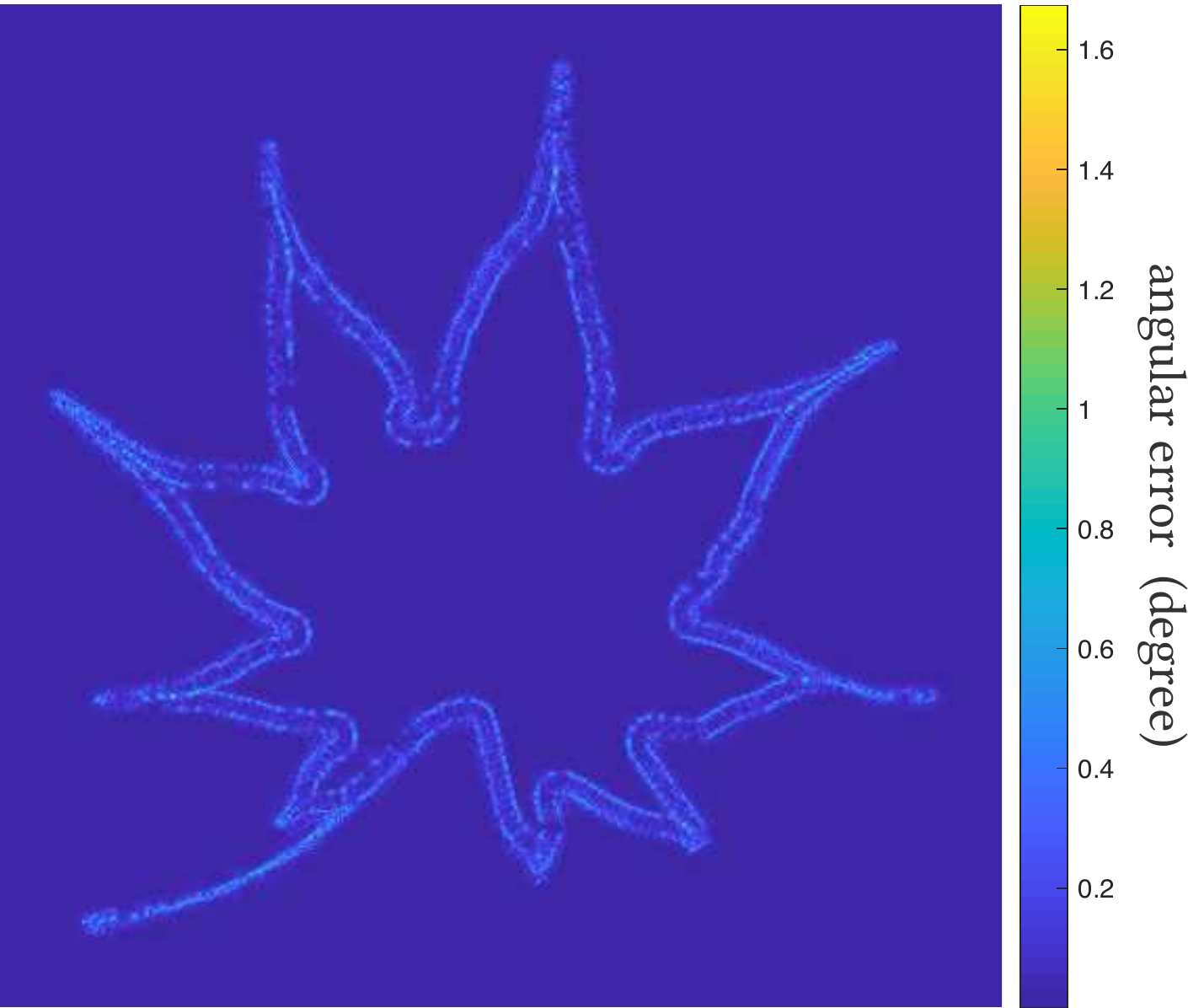}
		\caption{Scene: \figsym~\ref{figure: leaf}}
	\end{subfigure}
	\caption{\edit{\textbf{Error in the Poisson optimization.} The figure shows the absolute difference between the tilts caused by the phase function $\bphi$ and the desired tilts in \figsym~\ref{figure: sunny} and in \figsym~\ref{figure: leaf}. The average error near occluding boundaries are $0.09^\circ$ in (a) and $0.12^\circ$ in (b).
	}}
	\label{fig:poisson_err}
\end{figure}

\edit{
\subsection{Accuracy of Poisson optimization}  
Since the displayed phase pattern is obtained by solving a Poisson optimization problem over the desired field of cone tilts, it is not guaranteed that the phase function would tilt the light cones exactly by the desired amounts.
We empirically observe that this error is very small.
For example, in \figsym~\ref{figure: sunny}, the average angular error over the entire SLM is $0.001^\circ$ and is $0.09^\circ$ (or $4 \%$ of the largest SLM tilting angle) near the occlusion boundaries.
\figsym~\ref{fig:poisson_err} visualizes this error  for the scenes in Figs.\ \ref{figure: sunny} and \ref{figure: leaf}.
As can be seen, the errors are concentrated at the inner-most occluded pixels, where we have large changes in the tilting angles.
}

\subsection{Artifacts}

The captured results also  shows many of the artifacts in \conetilt displays.
We can see the dark halo in \figsym~\ref{figure: chess}b around the rook and \figsym~\ref{figure: leaf}b around the leaf (shown in the blue inset).
Note that when removing directly occluded background, the multifocal display also suffers from dark halo.
The \conetilt display also fails to prevent light leakage when two occluding boundaries are too close, as can be seen in \figsym~\ref{figure: leaf}b at the narrow breaking of the leaf (yellow inset) and in the railings in \figsym~\ref{figure: bridge}c (cyan insets).
We point out that there is some light leakage at the tips of leaf and the stem (pink inset).
This is due to the smoothness constraint we apply when solving the phase function. 
For example, pixels at the upper part of the stem need to the tilted upward, whereas the bottom part needs to be tilted downward; this causes the center portion of the stem to be un-tilted. 
The artifact can be removed by removing these background pixels, at a cost of increasing dark halo.

\begin{figure}[!ttt]
\centering
\includegraphics[width=0.475\textwidth]{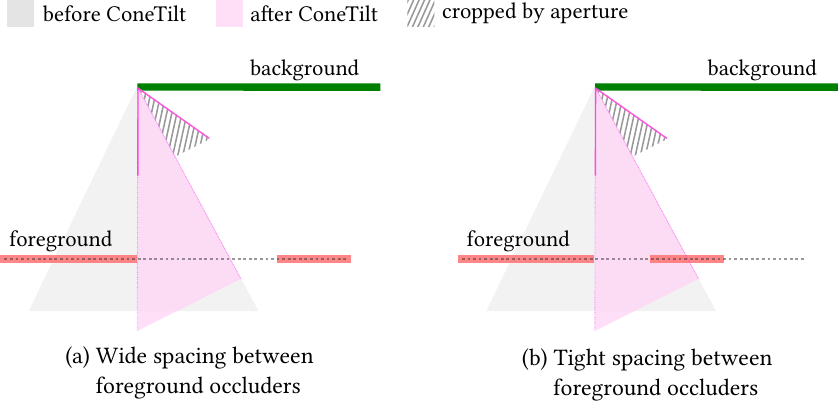}
\caption{\edit{\textbf{Infeasible \conetilt solutions for complex occlusions.} Tilting of light cones is insufficient for handling scenes with complex occluding shapes. A common scenario for such failure is when there are two occluding surfaces in close proximity as in (b). Here,  all tilts of the light cone from the background pixel leads to some intersection with the foreground occluders and hence, there is no feasible tilt that can avoid light leakage and its associated artifacts.}}
\label{figure: complex}
\end{figure}

\subsection{Inability to Handle Complex Occlusion Patterns} 

\conetilt displays tilt entire light cones to mimic the effect of occlusion.
While avoiding the loss of spatial resolution, this idea does not extend beyond simple occlusion scenarios where the occluding contours are smooth and well separated.
For example, if the front focal plane has two occluding contours in close proximity, \edit{as in \figsym~\ref{figure: complex}}, \conetilt would be insufficient to produce the occlusion cue.
For such  a scenario, we will need to ``trim'' the light cone, an operation that is beyond the simple tilt operation that we implement.

The minimum distance between two occluding contours on a focal plane is the size of the light cone on the front focal plane.
From \equsym~\eqref{eq: w},  we have
\begin{equation}
\mbox{min distance} =  \frac{2 d^2 u_m}{\delta} \left| \frac{1}{z_o} \, {-} \, \frac{1}{z_i}\right| \mbox{ display pixels,}
\label{eq: min distance}
\end{equation}
where $z_o$ and $z_i$ is the depth of the focal planes, and $\delta$ is the pixel pitch of the display pixels.
On our prototype, when the front and the back focal planes are separated by 4 diopters, the minimum distance between two occluders on the front focal plane can be $36$ pixels. 
Note that \equsym~\eqref{eq: min distance} decreases quadratically in $d$, whereas the eye box only decreases linearly in $d$. 
This provides an advantageous trade-off between the minimum distance and the size of the eye box.
Specifically, we can allow much closer occluding contours if we are willing to slightly reduce the size of the eye box.

\subsection{Limitations Due to the Phase SLM}
\label{sec: slm limitation}
In addition to the limited capability to tilt light, using a phase SLM induce the following limitations on a \conetilt display.
\paragraph{Chromatic Aberration.} 
Since the phase of the light depends on its wavelength, the phase function is color-dependent.
To create a typical RGB display, we can use time-multiplexing and show each of the phase functions designed for each color sequentially.
To alleviate the chromatic aberration caused by polychromatic light, the phase functions need to be smooth.
Thereby, in the optimization problem~\eqref{eq: opt} we use the $\ell_2$-regularization to find a smooth solution.
Nevertheless, since the phase SLM is attached to the display panel, the chromatic aberrations will only appear in the defocused regions, \ie, on an out-of-focus content that has been tilted.

\paragraph{Phase Wrapping Artifacts.}
Since most phase SLMs can only achieve a phase delay of $2\pi$, the phase function will be wrapped multiple times across the entire display. 
Due to the dramatic change in phase values, the wrapping creates dark seams in the images we see.  
While using smooth phase functions helps alleviate the problem, in our experience, the most effective solution is to add a global phase offset and rapidly change its value within the exposure time of a frame.  
Changing the offset shifts the dark seams without affecting the content, thereby it effectively smooths the dark seams.

\paragraph{Diffraction Efficiency.}
\edit{
The limited range of phase delay and the discretization of phase SLMs also results in low diffraction efficiency.
When implementing large tilts, the phase functions will be very close to the Nyquist limit.
Specifically, the phase functions have large slopes, which, due to phase wrapping, go from $-\pi$ to $\pi$ rapidly and repeatedly --- much like a grating.
This grating-like phase function not only tilts the light cone along the desired direction and angle but also at integer multiples of the desired angle.
Thus, the desired angle receives less light. 
This can be seen in \figsym~\ref{fig:psf_aperture}, where the captured image at $ \Delta u = 1.5 u_m$ is dimmer than other images at smaller tilts.
We refer to \cite{laude1998twisted} for detailed explanations.
}

\paragraph{Refresh Rate.}
Ideally, each focal plane should be paired with its own phase function. 
However, typical phase SLMs have a refresh rate of $60$ Hz and limits the number of phase functions we can display within a frame. 
As mentioned earlier in \secsym \ref{sec: display}, we work around this limitation using a simple decomposition of a scene into a foreground and background, each comprising of a phase functions and its corresponding intensity and depth maps.
Nonetheless, a faster SLM would be invaluable in handling complex scene configurations and enabling color displays where each color channel will likely need its own phase patterns tuned to its specific wavelength.

Finally, we note that \conetilt as an operator need not be implemented on phase SLMs.
We can use other technologies that can steer light locally, like the micro-prism proposed by~\citeN{smith2006agile}, which enables $\pm 7^\circ$  tilts.
This can improve the size of the eye box and the field of view of the display significantly.

\subsection{Miniaturization}%

\begin{figure}[!ttt]
	\centering
	\includegraphics[width=\linewidth]{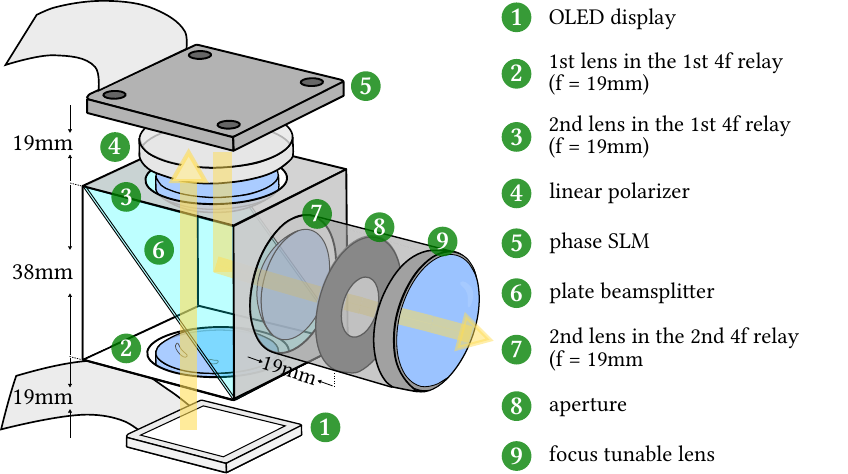}
	\vspace{-6mm}
	\caption{\edit{\textbf{Hypothesized design for miniaturization.} This figure illustrates one possible miniaturization of \conetilt displays. The design uses customized parts to hold three lenses within the cage cube, and it uses a high-speed OLED display panel whose angular range is controlled by the microlenses attached to the pixels. A field lens can be attached to the OLED or the SLM or implemented by the SLM.}}
	\label{fig:miniature}
\end{figure}

\edit{
Practical adoption of \conetilt displays requires a significant reduction in the footprint of the device.
Much of the bulk of our lab prototype is contributed by the off-the-shelf components to build a high-speed display with pixels that emit a light cone of specific angular range.
We can, hence, avoid this bulk and achieve a miniaturized prototype by using customized components.
\figsym~\ref{fig:miniature} illustrates such a hypothesized design, using the prototype of \citeN{matsuda2017focal} as a starting point.
The key component of this display is an OLED or LCD panel that has a refresh rate sufficiently high enough to display the desired number of depth planes and frame rate.
Further, the angular range of light emitted by each pixel needs to be matched to that of the aperture of the eyepiece; in principle, this can be realized during the manufacture process by adding a microlens onto each display pixel, similar to the method used in image sensors.
A $4f$ relay is used to colocate the display panel and the phase SLM, and the same relay is used to redirect light to the eyepiece by the beamsplitter.
This design also utilizes customized housings to hold all of the components.
Note that by adopting a transparent phase SLM, we can eliminate the need of a beamsplitter and further reduce the bulk of the display.
}

\section{Conclusion} \label{sec:discuss}

This paper proposes a simple but effective technology for displaying immersive virtual scenes on multifocal displays. 
The proposed display enables occlusion cues between focal planes of a multifocal display.
This has the dual effect of effect of enhancing the range of perceptual cues that the display can satisfy as well as reducing the loss of contrast due to leakage of defocus blur.
While our current prototype is bulky and limited by the capability of our phase SLM, the proposed \conetilt operator can be easily  incorporated into existing multifocal displays, while benefiting from the rapidly-evolving light modulation technologies.
Hence, we believe that the technology proposed in the paper will spur innovation in  virtual and augmented reality systems as well as  traditional light-field displays.

\begin{acks}
	The authors acknowledge support via the NSF CAREER grant CCF1652569.
\end{acks}

\bibliographystyle{ACM-Reference-Format}
\bibliography{main}

\end{document}